\newif\ifusebblmode\usebblmodefalse 
\newif\ifusetabularsource\usetabularsourcetrue

\ifusebblmode
  \documentclass[
  english,
onecolumn, 11pt,authoryear]{article}
\else
\documentclass[
  english,onecolumn, 11pt]{article}
\fi
\usepackage[T1]{fontenc}  
\usepackage[utf8]{inputenc}
\usepackage{geometry}
\usepackage[table]{xcolor}
\usepackage{color}
\usepackage{babel} 
\usepackage{array}
\usepackage{refstyle}
\usepackage{enumitem}
\usepackage{amsmath}
\usepackage{amsthm}
\usepackage{amssymb}
\usepackage{graphicx}
\usepackage{xspace}

\usepackage[authoryear]{natbib}

\definecolor{hyperlinkcolor}{rgb}{0, .25, 0}

\usepackage[unicode=true,pdfusetitle,
 bookmarks=true,bookmarksnumbered=false,bookmarksopen=false,
 breaklinks=true,pdfborder={0 0 0},pdfborderstyle={},backref=false,colorlinks=true]
 {hyperref}
\hypersetup{
 allcolors=hyperlinkcolor}

\usepackage{fancyhdr}

\usepackage{pdfcomment} 
 
\usepackage{paralist}

\usepackage[font=small,aboveskip=1pt,labelfont=bf,labelsep=period,justification=raggedright,singlelinecheck=off,figurename=Figure]{caption}

\newref{fig}{refcmd={\hyperref[#1]{Figure \ref{#1}}}}
\newref{tab}{refcmd={\hyperref[#1]{Table \ref{#1}}}}
\newref{eq}{refcmd={\hyperref[#1]{Eq.\ (\ref{#1})}}}
\newref{sec}{refcmd={\hyperref[#1]{Section \ref{#1}}}}
\newref{appsec}{refcmd={\hyperref[#1]{AppendiX \ref{#1}}}}
\newref{apptab}{refcmd={\hyperref[#1]{AppendiX Table \ref{#1}}}}
\newref{appfig}{refcmd={\hyperref[#1]{AppendiX Figure \ref{#1}}}}
\newref{prop}{refcmd={\hyperref[#1]{Proposition \ref{#1}}}}

\usepackage{afterpage} 

\usepackage{amsthm}
\usepackage{amssymb}
\usepackage{amsmath} 

\providecommand{\casename}{Case}
\providecommand{\propositionname}{Proposition}
\providecommand{\remarkname}{Remark}
\providecommand{\theoremname}{Theorem}
\theoremstyle{plain}

\theoremstyle{plain}

\newlist{casenv}{enumerate}{4}
\setlist[casenv]{leftmargin=*,align=left,widest={iiii}}
\setlist[casenv,1]{label={{\itshape\ \casename} \arabic*.},ref=\arabic*}
\setlist[casenv,2]{label={{\itshape\ \casename} \roman*.},ref=\roman*}
\setlist[casenv,3]{label={{\itshape\ \casename\ \alph*.}},ref=\alph*}
\setlist[casenv,4]{label={{\itshape\ \casename} \arabic*.},ref=\arabic*}
\theoremstyle{remark}
\newtheorem*{rem*}{\protect\remarkname}

\usepackage{bbm} 

\usepackage{tikz} 
\usetikzlibrary{shapes,decorations,arrows,calc,arrows.meta,fit,positioning}
\tikzset{
    -Latex,auto,node distance =1 cm and 1 cm,semithick,
    state/.style ={ellipse, draw, minimum width = 0.7 cm},
    point/.style = {circle, draw, inner sep=0.04cm,fill,node contents={}},
    bidirected/.style={Latex-Latex,dashed},
    el/.style = {inner sep=2pt, align=left, sloped}
}

\usepackage{floatpag} 

\usepackage{booktabs}  
\usepackage{tabularx}  
\usepackage{pdflscape} 

\usepackage{minitoc}  

\usepackage{multirow}
\usepackage{relsize}
\newcommand{\ctNtabCols}{}
\newcommand{\ctFirstHeader}{}
\newcommand{\ctSubsequentHeaders}{}

\newcommand{\ctBody}{}

\newcommand{\ctCaption}{}

\newcommand{\ctStartTabular}{}
\newcommand\ctEndTabular{\end{tabular}}

\newcommand{\ctStartLongtable}{}

    \newcommand{\coefse}[1]{{\smaller\smaller (#1)}}
    \definecolor{cSignifOne}{rgb}{.92,1,.92}
    \definecolor{cSignifTwo}{rgb}{.78,1,.78}
    \definecolor{cSignifThree}{rgb}{.5,1,.5}
    \definecolor{cSignifThousandth}{rgb}{.6,1,0} 
    \newcommand{\thousym}{$^{\dagger}$}

    \newcommand{\wrapSigTenPercent}[1]{{\em #1}$^{+}$\cellcolor{cSignifOne}}
    \newcommand{\wrapSigFivePercent}[1]{{\bfseries\em #1}$^{}$\cellcolor{cSignifTwo}}
    \newcommand{\wrapSigOnePercent}[1]{{\bfseries #1}$^{\star}$\cellcolor{cSignifThree}}
    \newcommand{\wrapSigOneThousandth}[1]{{\bfseries #1}\thousym\cellcolor{cSignifThousandth}}

    \newcommand{\cpblColourLegend}{{\footnotesize Significance:~\begin{tabular}{cccc}
        \wrapSigOneThousandth{0.1\%}~~~~&
        \wrapSigOnePercent{1\%}~~~~&
        \wrapSigFivePercent{5\%}~~~~&
        \wrapSigTenPercent{10\%}
        \\ \end{tabular} }}
 \usepackage{threeparttable}
 
\begin{document}
\doparttoc
\dopartlof
\dopartlot
\renewcommand \thepart{}
\renewcommand \partname{}


\newcommand\atitletext{The econometrics of happiness:\\
Are we underestimating the returns to education and income? 
}



\title{The econometrics of happiness:\\
Are we underestimating the returns\\ to education and income? }


\date{   
  ~\\~\\ Published in the \href{https://doi.org/10.1016/j.jpubeco.2023.105052}{{\em Journal of Public Economics}, 2024}  \\
  \href{https://doi.org/10.1016/j.jpubeco.2023.105052}{{\sc doi:} 10.1016/j.jpubeco.2023.105052}  \\
     \href{http://alum.mit.edu/www/cpbl/publications/Barrington-Leigh-JPubE2024-focal-values.pdf}{Latest update available on my own site}.
}
\newcommand\acktext{Contact information at \href{https://alum.mit.edu/www/cpbl/address}{\tt http://alum.mit.edu/www/cpbl/address}.\\
I am grateful for discussion and comments from Fabian Lange, Kevin Lang, Andrew Oswald, Idrissa Ouili, Nadia DeLeon, two helpful referees, 
co-editor Keith Marzilli Ericson,
and audiences  in Vancouver, Chicago, Oxford, Cambridge, Montreal, and Winnipeg, among others. 
This work was supported by Canada's Social Sciences and Humanities Research Council (SSHRC) grant 435-2016-0531.
}
\author{C P Barrington-Leigh\footnote{    \acktext}
}

\newcommand\XX{
}

\maketitle

\begin{abstract}
This paper describes a fundamental and empirically conspicuous problem
inherent to surveys of human feelings and opinions in which subjective
responses are elicited on numerical scales. The paper also proposes
a solution. The problem is a tendency by some individuals --- particularly
those with low levels of education --- to simplify the response scale
by considering only a subset of possible responses such as the lowest,
middle, and highest. In principle, this ``focal value rounding''
(FVR) behavior renders invalid even the weak ordinality assumption
often used in analysis of such data. With ``happiness'' or life
satisfaction data as an example, descriptive methods and a multinomial
logit model both show that the effect is large and that education
and, to a lesser extent, income level are predictors of FVR behavior.
 A model simultaneously accounting for the underlying wellbeing and
for the degree of FVR is able to estimate the latent subjective wellbeing,
i.e.~the counterfactual full-scale responses for all respondents,
the biases associated with traditional estimates, and the fraction
of respondents who exhibit FVR. Addressing this problem helps to
resolve a longstanding puzzle in the life satisfaction literature,
namely that the returns to education, after adjusting for income,
appear to be small or negative. Due to the same econometric problem,
the marginal utility of income in a subjective wellbeing sense has
been consistently underestimated.\\

\noindent {\footnotesize {\sc keywords:} happiness; subjective wellbeing; life satisfaction;  modeling; 
income; education; welfare}
\end{abstract}







\newpage
\tableofcontents \newpage
\listoffigures
\listoftables
\newpage
\pagestyle{fancy}
\fancyhead[LO,RE]{Econometrics of happiness ({\em J. Public Econ}, 2024)}
\fancyhead[CO,CE]{}

\newcommand\SWB{\pdftooltip{\hyperlink{defSWB}{SWB}}{Subjective Well-Being (See page \pageref{def:SWB})}\xspace}
\newcommand\SWL{\hyperlink{defSWL}{\pdftooltip{LS}{Life Satisfaction  (See page \pageref{def:SWL})}}\xspace}
\newcommand\FVR{\pdftooltip{\hyperlink{defFVR}{FVR}}{Focal Value Rounding (See page \pageref{def:FVR})}\xspace}
\newcommand\FVRI{\pdftooltip{\hyperlink{defFVR}{FVR}}{Focal Value Rounding Index (See page \pageref{def:FVRI})}\xspace}

\section{Introduction\label{sec:introduction}}

Now firmly entrenched in the economics literature, in national statistical
agency data collection, and in the dialogue about progress and wellbeing,
survey-based subjective evaluations of life\footnote{The life satisfaction question and close cousins such as the Cantril
Ladder question are posed in numerous national and international social
surveys, both cross-sectional and panel. The U.K. Treasury's Green
Book includes instructions for how to use compensating differentials,
estimated from life satisfaction, to carry out cost/benefit calculations
for central government \citep{UK-Treasury-2021-GreenBook-supplement-wellbeing,MacLennan-Stead-Rowlat-UK-Treasury-2021-life-satisfaction-approach}.
 } are the basis for  estimating welfare benefits and costs of everything
from inflation and unemployment, to air pollution and being married
\citep[e.g.,][]{Blanchflower-et-al-JMCB2014-unemployment-inflation,Levinson-JPubE2012-happiness-pollution,Stutzer-Frey-JSE2006-marriage-causality-happiness}.
Estimates of the psychological benefit of increased income, using
this approach, are five decades old, and those evaluating the net
individual return of additional education have been carried out for
at least three decades. In terms of optimally allocating human resources,
not much could be more central than knowing the marginal utility of
income and of education.

\subsection{Responding to life evaluation questions}

However, the coherence and value of subjective evaluations of life
rely on a series of considerable cognitive tasks to be performed in
short order by the respondent. When asked,\footnote{Typically, cognitive evaluations of life consist of a single, subjective,
quantitative question like this one, and responses are used directly
as a cardinal or ordinal proxy for welfare, i.e., ``experienced''
utility \citep{Easterlin-1974}.} \emph{``Overall, how satisfied are you with life as a whole these
days,  measured on a scale of 0 to 10?''} a respondent must in some
sense\label{sec:list-of-cognitive-steps}    \begin{inparaenum}[(i)] 
\item conceive of the domains, expectations, aspirations or other criteria salient to her sense of experienced life quality or satisfaction; 
\item assemble evidence pertaining to each ideal, such as recent affective (emotional) states, significant events, and objective outcomes; 
\item appropriately weight and aggregate this evidence according to its importance to overall life quality, and 
\item project the result onto  the  discrete numerical scale specified in the question. 
\end{inparaenum}

This is without doubt a tall order, and any embrace of \textbf{subjective
wellbeing} (\hypertarget{defSWB}{\textbf{SWB}})\label{def:SWB} data, and especially the headline measure
of %
\textbf{life satisfaction} (\hypertarget{defSWL}{\textbf{\SWL}}),\label{def:SWL} rests on their remarkable
reproducibility and apparent cardinal comparability, possibly along
with the principle that any objective indicator of experienced wellbeing
must ultimately be accountable to a subjective one. While various
studies have sought, with limited success, to find differences in
interpretation of the \SWL question or norms of expression across cultures
and languages \citep{Helliwell-Barrington-Leigh-Harris-Huang-2010,Exton-Smith-Vandendriessche-OECD2015,Lau-Cummins-Mcpherson-SIR2005-cultural-bias-SWB,Clark-Etile-Postel-Vinay-Senik-VanderStraeten-EJ2005-latent-class-coefficients-SWB},
an important fact is that, uniformly across cultures, responding to
the question is cognitively demanding. This paper focuses specifically
on the consequences of an apparent heterogeneity across respondents
in their ease with the final, quantitative step in the process outlined
above.

The crux is that people with less facility with numbers may simplify
the numerical response scale for themselves. In particular, the evidence
below shows that some respondents restrict the set of numerical options
under consideration to a three-point scale consisting of the bottom,
middle, and top options, rather than the full set offered. This can
be expected to introduce complex biases in mean life satisfaction
and in estimated marginal effects on life satisfaction, in particular
with respect to education and other correlates of numerical literacy
itself.

I present evidence of the prevalence and quantitative significance
of this problem, with implications for the interpretation and analysis
of all numerical, subjective response scales. The language and empirical
examples all focus on the case of single-item \SWB questions, mostly
\SWL \citep{Cheung-Lucas-QoLR2014-single-item-SWL-vs-SWLS-validity},
which underlie the field of the ``economics of happiness''. While
most empirical studies make use of a cardinal interpretation of the
response scale in the life satisfaction question, and at least an
ordinality assumption is universal,\footnote{A common finding is that models assuming cardinality give similar
results to those assuming only ordinality \citep{Ferrer-i-Carbonell-Frijters-EJ2004}.} the ``\textbf{focal value rounding}'' (\hypertarget{defFVR}{\textbf{FVR}})\label{def:FVR} behavior, described
above, introduces a conspicuous violation of the ordinality of response
options. Because a number of governments are gearing up to carry
out cost/benefit analyses using regressions of \SWL data for budgeting
and program evaluation \citep{Frijters-Clark-Krekel-Layard-BPP2020-Happy-Choice-SWB-as-goal-for-government,Frijters-Krekel-2021-SWB-policy-handbook,happiness-research-institute-2020-WALYs,Grimes-2021-chapter-budgeting-for-wellbeing,FinanceCanada-QoL-framework-2021-April,UK-Treasury-2021-GreenBook-supplement-wellbeing},
proper econometric accounting for \FVR may have practical importance.

In order to estimate the size of systematic biases associated with
widely used methods of inference based on \SWB reports, I present a
model which accounts for the \FVR phenomenon and which shows why biases
on estimates can be large or small and positive or negative. The model
also quantifies the fraction of respondents in a sample who have chosen
an alternate, simplified response scale, a value I call the \textbf{Focal
Value Rounding Index}, or \hypertarget{defFVRI}{\textbf{FVRI}}.\label{def:FVRI}

The rest of this paper proceeds as follows. The remainder of the Introduction
reviews some  stylized facts related to education and wellbeing in
the happiness literature, and mentions some points of history in the
development of \SWB survey questions like \SWL. Next, \secref{Motivational-evidence}
will convince the reader that there is a measurement problem with
quantitative, subjective scales like \SWL that is conspicuous, ubiquitous,
and strongly correlated with educational attainment  and that it
has a natural explanation  supported by the behavioral evidence.
Then \secref{Cognitive-model} presents the formal model in which
a mixture of high- and low-numeracy respondents treat the response
scale differently. \secref{simulation} validates the estimation and
identification approach using synthetic data and explores the complexity
of biases that can result from \FVR. \secref{empirical-estimates}
presents the main empirical estimates of the relationship between
education, income, and wellbeing, using a large social survey from
Canada. \secref{Applications} reexamines three previously published
studies, along with a ranking of U.S. states, as applications to investigate
the extent of bias in existing published literature as well as in
popular happiness rankings. In these empirical applications, previously
anomalous but reproducible findings include evidence that a disadvantaged
population reports high life satisfaction, and that the return to
extra years of education after primary school are negative, especially
when conditioned on income. These surprising findings are overturned
when taking into account focal response behavior. A summary and a
perspective on future directions are in \secref{Conclusion}.

\subsection{Effects of education and income on subjective wellbeing\label{sec:education-and-income-literature}}

Education may be expected to confer welfare benefits not just through
higher income but also through better health behaviors and enhanced
social capital of various forms with intrinsic benefit \citep[e.g.,][]{Helliwell-Putnam-EEJ2007,Powdthavee-Lekfuangfu-Wooden-JBEE2015-education-SWL-Australia-HILDA},
as well as through some kind of psychological capital which captures
intrinsic benefits of learning or knowledge, or which complements
other consumption (for instance, possibly literature, fine art, or
the night sky). However, among the more surprising stylized facts
in the economics of happiness is that formal education does not help
much to explain \SWL once income\footnote{Studies routinely control for current income rather than wealth when
modeling \SWL. Current income may be an especially poor proxy for lifetime
income in this context because choosing to pursue extra education
entails a trade-off between short-term income and future income. This
leaves educational attainment as a positive proxy for unmeasured future
income expectations, making the low coefficients measured on education
even more surprising.} is accounted for \citep[e.g., ][]{Layard-2011-happiness-lessons-new-science,Frijters-Krekel-2021-SWB-policy-handbook}.\footnote{This generalization hides considerable variation in the literature.
Since the various channels and directions of influence are not easily
identified, and the relationship may not even be monotonic \citep{Stutzer-JEBO2004},
estimates vary from slightly positive to substantially negative overall
effects of having extra education.}

Similarly, although the literature on the importance of income and
income growth on \SWL is enormous and involves a large potential role
of consumption externalities \citep{Barrington-Leigh-EQLWBR2014-consumption-externalities},
one may summarize the findings by saying that income has been found
to be a weak predictor of \SWL in comparison to other, less market-mediated
parts of life \citep[e.g.,][]{Blanchflower-Oswald-JPubE2004,Hamilton-Helliwell-Woolcock-NBER2016-social-capital-wealth,Layard-2011-happiness-lessons-new-science,Frijters-Krekel-2021-SWB-policy-handbook}.\footnote{That is, the large compensating differentials found for having positive
social relationships \citep[trust, engagement, meaningful work, friendships, intimate relationships, etc; see for example][]{Powdthavee-JSE2008-valuing-social-relationships,Helliwell-Barrington-Leigh-2010-social-capital-worth,Helliwell-Putnam-PTRSL2004}
reflect a small denominator, i.e.~the value of income for increasing
life satisfaction. }

This paper does not aim to explore all the reasons for this well-established
evidence about quality of life from subjective response data. Instead,
it characterizes a measurement error in which those with lower education
and, as a proxy, those with lower income, may be more likely to under-utilize
the \SWL response options in such a way that tends to bias their reported
life satisfaction. In general, resulting biases on marginal effects
could exist in either direction, but as described below they are more
likely to be downward, meaning that they may go some way to explaining
the education anomaly and to revise upward, if modestly, the estimated
importance of income for supporting \SWB.

\subsection{Evolution of precision in subjective, quantitative reports}

The history of survey questions on subjective assessments mirrors
in part technological norms. Early innovators in monitoring \SWB in
social and household surveys tended to use a four point or five point
scale, typically with Likert-style verbal response options. In such
questions, the numbers were not meant as cues for the respondent.
In some populations, most respondents chose one of the top two options,
limiting the variation, or precision. As limitations of paper survey
media have been erased by the adoption of computer aided interviews,
the resolution of these subjective scales has expanded. However, with
more than five or seven response options, verbal cues are typically
not provided except for the highest and lowest response options. Responses
instead become numerical. For instance, after many years of asking
\SWL questions with a variety of scales, Statistics Canada settled
over a decade ago on a particular wording with an 11 point scale.\footnote{However, \citet{Conti-Pudney-RES2011-SWB-survey-design-BHPS-distribution}
describe an evolution in the opposite direction, away from unlabeled
response options, in 1992 in the British Household Panel Survey.}

The \citet{OECD-2013-guidelines-measuring-SWB} has also developed
recommendations for standardizing the way such questions are asked
by all national statistical agencies. The \emph{de facto} standard
for \SWL now is an 11-point scaling, from 0 to 10, with the lower extreme
meaning, for example, ``not at all satisfied'', the upper signifying
``completely satisfied'', and the interpretation of the remaining
values left up to the respondent.

\label{sec:processing-and-expressive-capacities}

An older literature sought to determine the optimal number of response
options in survey questions with verbal cues for each option. For
instance, it may be that in an oral interview, i.e., with no visual
cues, four or five responses are the maximum that can be handled without
confusion or overload \citep{Bradburn-Norman-Sudman-Wansink-2004-questionnaire-design}.

When the scale is explicitly numeric, as with modern \SWL measures,
there also arises a trade-off between the cognitive load imposed by
a scale and the precision it allows. From the respondent's point
of view, this trade-off is between the opportunity for self-expression
and the cost of cognitive processing. The survey designer wishes to
allow for precise responses in order to capture variability among
respondents and over time, while not demanding too much. Overburdening
would result, at best, in the respondent not fully optimizing her
answer or not properly interpreting or using the given range of responses
\citep{OECD-2013-guidelines-measuring-SWB}. Various studies on
this balance have tended to favor 11-point quantitative scales over
coarser option sets (e.g. 7-point scales) as well as over nearly continuous
options \citep{Alwin-SMR1997-survey-question-scales,Kroh-DRAFT2006,Saris-vanWijk-Scherpenzeel-SIR1998,OECD-2013-guidelines-measuring-SWB,Weng-EPM2004-Likert-labels-and-number-of-options}.\footnote{Interestingly, government surveys in the USA have tended to stick
with 3 or 4-point scales for \SWB questions.}

\section{Descriptive evidence\label{sec:Motivational-evidence}}

A small number of studies have remarked in some way on the use of
focal values, but without a full account or explanation.\footnote{There is also a psychometrics literature which refers to the tendency
towards top and bottom value responses as ``extreme response style''
and tendency towards the central value as ``moderate response style''
\citep{Hamamura-Heine-Paulhus-PID2007-cultural-differences-response-styles,Khorramdel-vonDavier-Pokropek-BJMSP2019-extreme-response-styles-model}.
That literature, known in psychology as ``item response theory'',
is motivated by an interest in personality type, as classified by
responses to a battery of Likert questions with all-verbal response
options. These studies have not considered cognitive ability as an
explanatory factor. \citet{Giustinelli-Manski-Molinari-JEconometrics2020-rounding}
also study rounding of reported quantitative beliefs which relies
on observing multiple questions for each respondent.} \citet{Dolan-Layard-Metcalfe-ONS2011-SWB-public-policy} mention
that \SWL ratings in one study are positively associated with life
circumstances as one would expect, except at the top of the scale,
where ``those rating their life satisfaction as `ten out of ten'
are older, have less income and less education than those whose life
satisfaction is nine out of ten''. They speculate a reason unrelated
to cognitive limitations for this observation but declare that ``This
issue warrants further research''. \citet{Conti-Pudney-RES2011-SWB-survey-design-BHPS-distribution}
describe focal value behavior as a response to the existence of verbal
cues, present on only three out of seven response options.  \citet{Landua-SIR1992-panel-focal-values}
analyses response transition probabilities in the German Socio-Economic
Panel, and \citet{Frick-Goebel-Schechtman-Wagner-Yitzhaki-SMR2006-SOEP-attrition-endpoints}
confirm his report that respondents have a tendency to move away from
the endpoints over time. In fact, this could be driven largely by
the \FVR behavior diminishing as panel participants, especially those
with low numeracy, gain familiarity and comfort with the scale.

\subsection{Educational attainment}

Simply inspecting their \SWL distributions, stratified by education
level, might have led these authors to the hypothesis developed in
this paper. For illustrative purposes I appeal to one cycle from the
Canadian Community Health Survey (CCHS), a large annual cross-section
which includes an 11-point life satisfaction question as well as educational
attainment.\footnote{In a repeated cross-section, most respondents are facing the \SWL question
for the first time. In the CCHS, education is recorded in four categories:
less than high school graduate, high school graduate, some post-secondary
training, and completed post-secondary training, but relatively few
respondents report the third category, so I combine the top two. In
the 2017-2018 wave, out of a total of 113289 respondents, 97604 reported
their age as at least 25 years, and 93043 of those answered the \SWL,
educational attainment, and income questions.

} Conditioning \SWB responses on educational attainment reveals a striking
feature (\figref{CCHS-17-18:SWL-by-education}).  The relative
frequencies of each focal value (0, 5, and 10) decrease with increasing
education level. While the lowest education category shows four peaks,
the distribution of responses in the highest education category features
what would be a unimodal distribution around SWL=8, except for a slight
enhancement at SWL=0.   In addition to \figref{CCHS-17-18:SWL-by-education},
several other lines of evidence support the interpretation that scale
simplification is a specific response to cognitive challenge, a model
to be formalized in \secref{Cognitive-model}. 

\subsection{Numeracy}

\begin{figure*}
\begin{centering}
\par\end{centering}
\begin{centering}
\includegraphics[width=\textwidth]{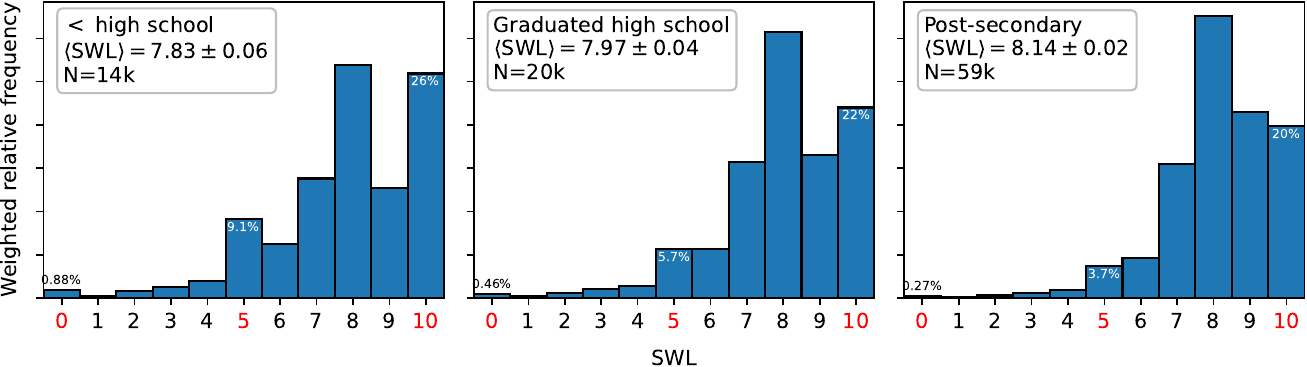}\\
\par\end{centering}
\caption[CCHS SWL responses versus education]{CCHS: \SWL responses versus education for those at least 25 years
old. Education is coded into three categories. Response fractions
within each education category are shown for focal value responses.
Use of focal values decreases with increasing education. Mean \SWL
in each group is shown with 95\% confidence range. Histograms and
means are population estimates, using sampling weights provided by
Statistics Canada. \label{fig:CCHS-17-18:SWL-by-education}}
\end{figure*}

Educational attainment is a widely-available characteristic in social
survey data, but may capture attributes which are relevant to latent
wellbeing or to reporting functions, but which are different from
cognitive ability with numbers. To support the numeracy interpretation
of \FVR, \href{https://alum.mit.edu/www/cpbl/publications/Barrington-Leigh-JPubE2024-appendix.pdf}{Appendix Figure F.1} makes use of the 2018 Programme for
International Student Assessment (PISA) survey of 15 year old enrolled
school students in 72 countries. This survey includes a measure of
mathematical ability, along with 11-point life satisfaction. Separating
respondents according to an overall math score,\footnote{This math score is a ``plausible value'' of the individual's underlying
latent ability, appropriate for modeling relationships such as this
one; see \citet{Wu-SEE2005-plausible-values-PISA}.} the same feature as in \figref{CCHS-17-18:SWL-by-education} is observed:
the relative frequency of each focal value response \emph{decreases}
with math proficiency, as does the mean \SWL response. This is true
of the global distribution, as well as for individual countries such
as the USA (see \href{https://alum.mit.edu/www/cpbl/publications/Barrington-Leigh-JPubE2024-appendix.pdf}{Appendix Figure F.1}).

\subsection{Difficulty responding}

Another indication that the apparent tendency to simplify the response
scale has to do with the difficulty of answering the question, as
it is posed, comes from noticing that respondents with less education
are more likely to refuse to answer the \SWL question at all. \href{https://alum.mit.edu/www/cpbl/publications/Barrington-Leigh-JPubE2024-appendix.pdf}{Appendix Table F.1}
shows that response rates to the \SWL question, although close to
100\%, are strictly increasing with educational attainment.

\subsection{Unordered choice model}

The existence of \FVR behavior implies that \SWB response scales cannot
safely be assumed to be ordinal. For example, those with lower education
may, all else equal, experience lower life satisfaction but be systematically
inclined to report a higher value due to rounding up from a 3 or 4
to 5, or from 8 or 9 to 10. It is possible, therefore, that on average
those reporting 9 could be happier than those reporting 10.

Traditional methods used in econometric inference from \SWL --- such
as OLS, ordered logit, ordered probit, and related time series and
instrumented analogues --- leverage strong assumptions about the symmetry
of effects of explanatory variables on each step of the response scale,
as well as assuming cardinality or at least ordinality among response
values. Those models are therefore not flexible enough to account
for the heterogeneous influence of predictors like education on focal
and non-focal response values.\footnote{Note that, unlike OLS, ordered logit and ordered probit naturally
account for multi-peaked distributions such as is shown in \figref{CCHS-17-18:SWL-by-education}.
That is due to the flexibility given by the cut points in those models,
which can squeeze together or stretch apart in order to decrease or
increase (respectively) an option's predicted response probability.}

An alternative approach is to relax the ordinality assumption for
response options, and model the probability of each response independently,
subject only to the constraint that the probabilities add up to one.
The  multinomial (polytomous) logit model\footnote{The multinomial logit model, in its latent variable formulation, consists
of a system of equations generating scores $Y_{i,j}^{*}$ for each
individual $i$ and response option $j\in\{0\dots10\}$ as $Y_{i,j}^{\ast}=\boldsymbol{\beta}_{j}\cdot\mathbf{x}_{i}+\varepsilon_{j}\,$
where $\varepsilon_{j}\sim\text{EV}_{1}(0,1),$ i.e., the error terms
have standard type-1 extreme value distributions. Then observation
probabilities are given by $P(Y_{i}=s_{j})=e^{\boldsymbol{\beta}_{j}\cdot\mathbf{x}_{i}}/\sum_{k\ne j}e^{\boldsymbol{\beta}_{k}\cdot\mathbf{x}_{i}}$
with one necessary normalization like $\beta_{0}=0.$ This model
is clearly also misspecified for \SWL data, since the formal \emph{independence
of irrelevant alternatives} (IIA) assumption is violated by the numbered
options of an \SWL question. The IIA requirement is frequently overlooked
in applications of the multinomial logit.} does this.

\begin{figure*}
\begin{centering}
\includegraphics[width=1\textwidth]{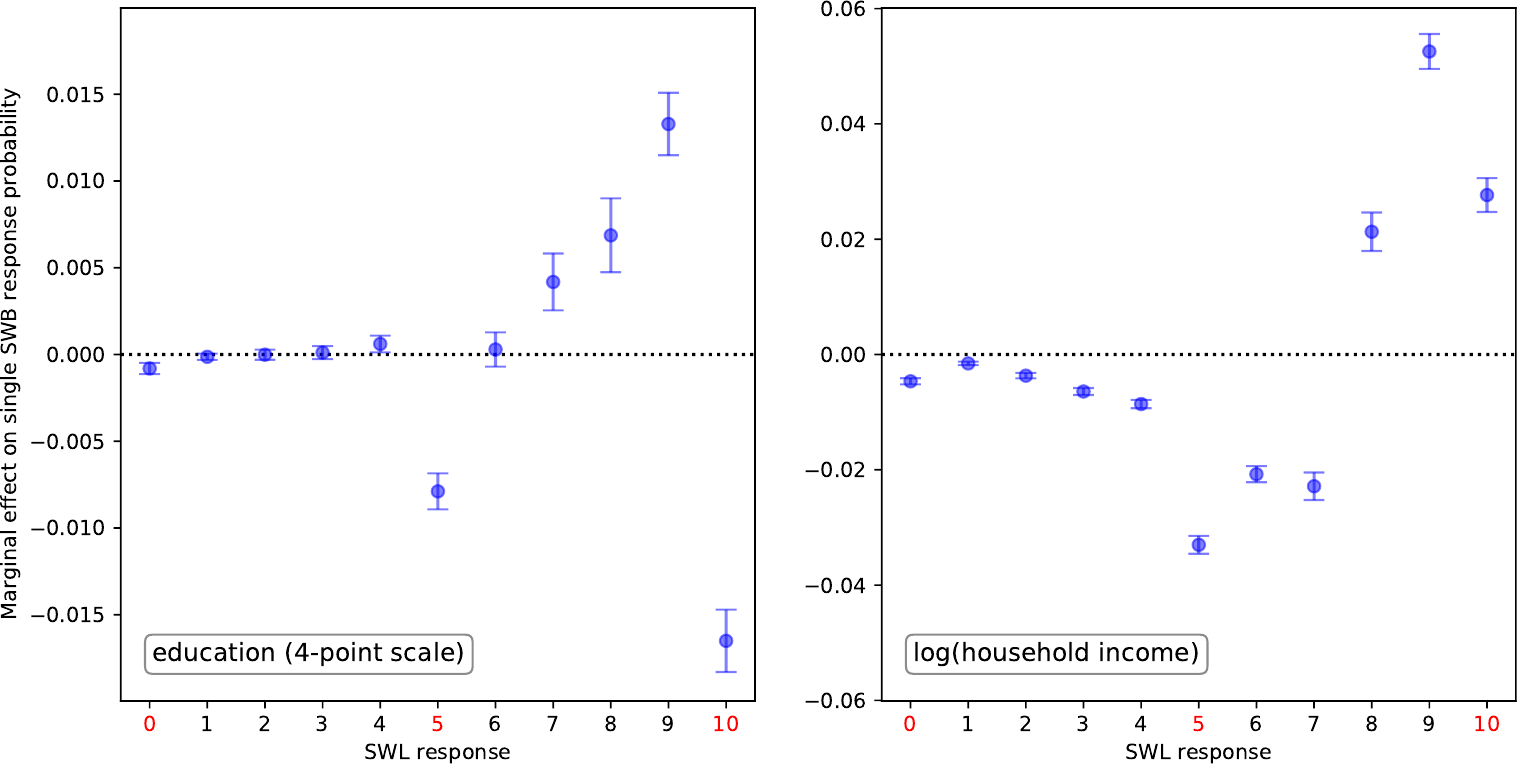}
\par\end{centering}

\centering{}\caption[Marginal effects of education and income on probabilities of each
response option]{ Multinomial logit estimate of individuals' probability of giving
each possible response to the life satisfaction question (CCHS data).
For educational attainment (quantified on a 1--4 scale) marginal effects
show a monotonic pattern with increasing response value, except for
the remarkable outliers at 0, 5, and 10. These indicate that education
and, simultaneously but to a lesser degree, income are significant
predictors of the tendency to use a simplified response scale. Error
bars denote 95\% confidence intervals.\label{fig:mlogit-marginal-effects-multivariate}}
\end{figure*}

\global\long\def\Pswl#1{P_{\text{SWL=#1}}}%

\figref{mlogit-marginal-effects-multivariate} shows marginal effects
of education and income on response probabilities of each of the 11
points in the \SWL scale, from a multinomial logit model using education,
logarithmic income, age, and age$^{2}$ as predictors for the sample
shown in \figref{CCHS-17-18:SWL-by-education}.  Under an ordinality
assumption, one might expect marginal effects to rise monotonically
with response value, since a better circumstance like education or
income should lead to an increase in the relative probability of response
$s+1$ as compared with response $s$. Indeed,  other than the focal
response values 0, 5, and 10, the marginal effect of one step higher
educational attainment (for instance, graduating from high school)
is weakly increasing in reported \SWL. By contrast, the effects on
the focal value responses are, with high statistical significance,
negative\footnote{For each explanatory variable, the sum of all marginal effects on
probabilities is zero by construction.} outliers far below what would be expected based on the pattern of
adjacent values. They show that more education significantly \emph{reduces
}the probabilities of each focal value response. Multinomial logit
estimations provide a diagnostic tool for detecting predictors of
focal value behaviour. However, the effect sizes are hard to interpret
because they are averages over the entire sample. For instance, the
education coefficient for \SWL=10 is an average effect over \emph{high}
types, for whom higher education increases the chance of reporting
10, and \emph{low} types, for whom higher education decreases that
chance. In order to separate those effects, a more structured mixture
model approach, described below, is required. 

\subsection{Precision and self-expression}

As a final piece  of empirical motivation for the modeling approach
developed below, I note that when excess precision is offered in
an \SWB scale, respondents appear to make a costly effort to choose
round numbers. Specifically, \href{https://alum.mit.edu/www/cpbl/publications/Barrington-Leigh-JPubE2024-appendix.pdf}{Appendix Figure F.2}
shows the distribution of responses from a computer-based \SWB survey
question framed on a 0--10 scale but with an available resolution of
0.1. There are clearly favored responses at every integer and half-integer
value. The response interface was a graphical slider which gave no
preference for any particular values. Thus, the prevalence of rounded
values indicates that \emph{extra effort} in the form of fine manual
control was exerted in order to leave the slider precisely on a half-
or whole-integer value. This can be interpreted as evidence of effort
to faithfully communicate a mental result, motivated by the drive
for self-expression \citep{Alwin-SMR1997-survey-question-scales,OECD-2013-guidelines-measuring-SWB}.\footnote{It also suggests that respondents have introspective knowledge of
their degree of precision in answering a numerical \SWB question ---
knowledge which is not typically elicited in surveys.}

\section{Cognitive mixture model\label{sec:Cognitive-model}\label{sec:theory}}

\global\long\def\Sstar{S^{\star}}%
\global\long\def\Nstar{N^{\star}}%
\global\long\def\Prob#1{P\left(#1\right)}%
\global\long\def\Prp{\Prob{\#1}}%
\global\long\def\Prg#1#2{\Prob{#1\mid\boldsymbol{x}#2}}%
\global\long\def\Prgz#1#2{\Prob{#1\mid\boldsymbol{z}#2}}%

Motivated by the evidence above, the enhanced use of focal values
can be interpreted as an indication that respondents have simplified
their cognitive task by coarsening the numerical scale. Because \FVR
behavior is inversely associated with education and math skills, I
focus on ``numeracy'' as one major influence on scale choice. 
The two-type mixture model below is based on the assumption that the
cognitive processes of respondents differ only in the execution of
step (iv) described in the second paragraph of \secref{introduction}.
That is, an internal representation of overall wellbeing exists in
a similar way across the two groups, who subsequently project that
assessment onto either the full scale (\emph{high numeracy} respondents)
or a subset consisting of the bottom, central, and top values (\emph{low
numeracy} respondents).

For each of the two types, latent wellbeing is mapped onto a discrete
response scale as in a standard, i.e.~canonical, ordered logit model.
That is, given a continuous, latent subjective assessment $S^{\star}$
modeled in terms of individual characteristics $\boldsymbol{x}$ as
$S^{\star}=\boldsymbol{x'\boldsymbol{\beta}_{s}}+\varepsilon$, the
cumulative probability of discrete responses $k$ is given by: 

\begin{align}
\Prg{s\mathcal{}}{}=\Prg{S^{\star}>\alpha_{k}}{} & =\Prob{\boldsymbol{x'\boldsymbol{\beta}_{S}}+\varepsilon>\alpha_{k}}\nonumber \\
 & =1-\Phi\left(\alpha_{k}-\boldsymbol{\text{x}'\boldsymbol{\beta}_{S}}\right)\label{eq:cumulativeOLogit}
\end{align}
where $\alpha_{k}$ are a sequence of threshold values $\alpha_{k}^{H}$
separating the full set of observed responses $\left\{ 0,1,\dots,10\right\} $,
or $\alpha_{k}^{L}$ for the focal subset $\left\{ 0,5,10\right\} $,
and $\Phi(\cdot)$ is the cumulative distribution function of the
unexplained portion $\varepsilon$ of $S^{\star}$. Use of the logistic
distribution for $\Phi(\cdot)$ makes this an ordered logit model.

The \emph{high} and \emph{low} alternative ordered logit outcomes
are combined using a simple dichotomous logit model. If $\boldsymbol{z}$
is a vector of individual characteristics, possibly overlapping with
$\boldsymbol{x}$, which serve as a measure of numeracy, then 
\begin{equation}
\Prgz{\text{high}}{}=\Phi\left(\boldsymbol{z'\boldsymbol{\beta}}_{N}\right)\label{eq:PhighLogit}
\end{equation}
There is no explicit consideration of costs and benefits to the respondent.\footnote{Conceptually, the individual benefits of using the full scale are
self-expression and performing one's best at fulfilling the purpose
of the survey; the costs are those of the cognitive calculation. For
more on the tradeoff between self-expressive capacity and cognitive
capacity in the design of response scales, see \citealt{Alwin-SMR1997-survey-question-scales,Kroh-DRAFT2006,Saris-vanWijk-Scherpenzeel-SIR1998,OECD-2013-guidelines-measuring-SWB}.
There is no evidence that respondents' value of time, which might
for instance vary with income, is also a major factor in scale choice
in the face of the inclination for self-expression. The value of time
likely affects the choice to participate in a survey, but once asked
the \SWL question, respondents answer it quickly, i.e., in a few seconds.
The indication in evidence shown above and below is that higher income
predicts higher resolution in responses, which seems contradictory
to a hypothesis of behavior driven by the market value of respondents'
time..}  

\newif\iftwocolumncpbl\twocolumncpbltrue
 Together, \eqref{cumulativeOLogit,PhighLogit} form
 a mixture model. The probability of observing response $k$ is
 \iftwocolumncpbl
     \begin{align}
  \Prg k{,\boldsymbol{z}}=&\Prgz{\text{high}}{}\Prg k{,\text{high}}  \nonumber \\
                            & +\left[1-\Prgz{\text{high}}{}\right]\Prg k{,\text{low}}\label{eq:one-line-probability}
\end{align}
\else
\begin{equation}
\Prg k{,\boldsymbol{z}}=\Prgz{\text{high}}{}\Prg k{,\text{high}}+\left[1-\Prgz{\text{high}}{}\right]\Prg k{,\text{low}}\label{eq:one-line-probability}
\end{equation}
\fi  
The model is similar to the ordinal-outcome ``finite mixture model''
of \citet{Boes-Winkelmann-ASA2006-ordered-response} except that here
the mixing probability is dependent on individual characteristics
\citep[see also][]{Everitt-Merette-JAS1990-mixture-model-ordered-logit,Everitt-SPL1988-mixture-model-ordered-logit,Uebersax-APM1999-mixture-model-ordinal}.
A more detailed account of the model is presented in \href{https://alum.mit.edu/www/cpbl/publications/Barrington-Leigh-JPubE2024-appendix.pdf}{Appendix A.}.

\subsection{Identification}

Are the parameters in this model point-identified in principle?\footnote{Point-identification, typically referred to simply as ``identification'',
is also called frequentist identification and is a frequentist concept.
In Bayesian estimation, as is used in the empirics to follow, parameters
are assumed to have distributions, not point values. Using a Bayesian
estimation method with a broad enough prior, alternate sets of values
which account for the data are simply reflected in multimodal (or
suitably broad) estimates of the parameters \citep{Lewbel-JEL2019-identification-meaning}.} Identification is a challenge because the same predictors may be
used to predict the latent numeracy variable and to predict the latent
wellbeing variable. As a result, one might fear that more than one
set of parameters could equally well explain observations for a given
sample.  Excluding the columns of $z$ from $x$ in \eqref{one-line-probability}
would overcome this problem. However, for an all-encompassing subjective
outcome such as latent wellbeing, it is safer to assume that everything
could be a determinant. More specifically, a particular interest
motivating this study is to assess the bias on estimates of the wellbeing
effect of education, and education is also the primary available predictor
of numeracy.

With stronger assumptions, an alternative strategy to the mixture
model may be able to identify parameters for latent \SWB by avoiding
\FVR altogether. One approach would be through thin set identification,
if respondents with some level of education were known never to exhibit
\FVR. One standard problem with this kind of identification is that
it relies on an assumption of a uniform effect of a covariate across
its support, as well as the absence of interaction effects with other
covariates. By contrast, the mixture model approach of \eqref{one-line-probability},
which leverages the entire sample, has the advantage of generalizability
to explicitly estimate interaction terms or other functional forms
to allow for non-uniform effects.

Another approach would be through selection on the dependent variable;
that is, by restricting the sample to the subset of high types who
did not respond with a focal value. Because no ``5''s are observed
in this group, it would consist of two subsamples: those with observed
$s\in\{1,2,3,4\}$ and those with $s\in\{6,7,8,9\}$. In fact, if
the symmetries required for this approach to be unbiased were believed,
then one could likely estimate coefficients for latent wellbeing using
binary models like logit and sample subsets of respondents who answered
one of only two consecutive, non-focal response options.

Returning to \eqref{one-line-probability}, within each of the two
ordered logit formulations nested in the model, identification of
the set of parameters (with no constant term) is standard. This still
leaves us with incomplete identification, in general, of the parameters
on variables common to $\boldsymbol{x}$ and $\boldsymbol{z}$. One
can imagine extreme distributions of \SWB, for instance all near 10,
in which \FVR only acts to convert 9s to 10s. In this case, discriminating
between the effect of common variables on latent wellbeing or \FVR
would not be possible, especially if the sign of coefficients is not
constrained. However, more typically, with a broader \SWB distribution,
\FVR will be distinguishable from effects on latent wellbeing. That
is, successful identification rests on having sufficient independent,
explainable variance in latent \SWB across low types in order that
there is also variation in their observed response. For instance,
if the latent wellbeing of low types is sufficiently spread out that
they sometimes round down and sometimes round up, then the influence
of education on \SWB, controlling for other influences, is separately
identified from the influence of education on the reporting function,
i.e., on the likelihood of being a low type. Put differently, identification
comes from the response of the observed distribution to changes in
numeracy, driven by some variable, being different from the response
of the observed distribution to changes in latent wellbeing, driven
by the same variable. This is assured when there is nontrivial variation
in the latent wellbeing of low types. This conceptual argument is
best corroborated quantitatively through simulation, which demonstrates,
in \secref{simulation}, that $\boldsymbol{\beta}_{S}$ and $\boldsymbol{\beta}_{N}$
are simultaneously recovered when estimating \eqref{one-line-probability}.

\subsection{Focal Value Rounding Index}

As shown below, net biases on some estimated moments and model coefficients
may be zero due to offsetting effects,  even when \FVR behavior is
prominent.  Therefore,  to express straightforwardly the magnitude
of the numeracy problem in a sample of respondents, another estimated
value is helpful. This is the Focal Value Rounding Index (\FVRI), which
is an estimate of the fraction of the population who restrict their
answer to a set of focal values --- i.e., the estimated fraction of
low types. This value is well identified whenever $\boldsymbol{\beta}_{N}$
is.

\subsection{Counterfactual SWL distribution}

The mixture model provides a posterior estimate of the latent \SWB
distribution, i.e., that which would have been reported had respondents
all used the full scale. This represents a ``correction'' to the
reported distribution of \SWB. This is a distribution of predicted,
counterfactual, discrete responses on the 0--10 scale, not an estimate
of the latent variable $S^{\star}.$ The next section demonstrates
through simulation that the model successfully recovers (identifies)
this counterfactual distribution, along with the \FVRI, means, and
coefficients.

\section{Model validation\label{sec:simulation}}

This section, supplemented by several appendices, reports on the use
of simulated data to validate the computational approach\footnote{Estimation of the mixture model was carried out using the no-U-turn
sampler (NUTS) variant of a Hamiltonian Monte Carlo (HMC) algorithm,
which is in turn a Markov Chain Monte Carlo (MCMC) method \citep{Stan2018,pystan2.19.1.1,Carpenter-Gelman-Hoffman-Lee-Goodrich-Betancourt-Brubaker-Guo-Li-Riddell-JournalStatSoftware2017-Stan}
and handles the non-concave objective well. \href{https://alum.mit.edu/www/cpbl/publications/Barrington-Leigh-JPubE2024-appendix.pdf}{Appendix B}
provides more detail on estimation, including analytic derivations
of the gradient and Hessian for a log likelihood approach.} and the model's ability to identify simultaneous influences of a
predictor, like education, on \FVR and on latent \SWB.  A large battery
of simulations demonstrates the complexity and scope of possible
biases, due to \FVR, in conventional estimates of \SWB means and of
marginal effects.\footnote{To reiterate, this bias is, conceptually, the difference between naively
estimated values and those which would be obtained in the counterfactual
case that all respondents had used the full numeric response scale,
i.e.~were of ``high numeracy'' type.  This counterfactual can
be perfectly calculated using synthetic data, but is of course unobservable
in traditional empirical data.}

\subsection{Synthetic data validation results\label{sec:Validation}}

Simulated \SWB data are generated by a model in which a scalar $z$
partly determines numeracy through \eqref{PhighLogit} while $z$
and a second scalar, $y$, partly determine the latent wellbeing $S^{\star}$
(thus $\boldsymbol{x}\equiv\begin{bmatrix}z\\
y
\end{bmatrix}$ in \eqref{cumulativeOLogit}). A non-zero correlation, parameterized
by $\chi$, may exist between $z$ and $y$. Conceptually, and for
comparison with the empirical estimates to follow, $z$ is meant to
represent education and $y$ represents other variables, such as income,
which are not direct measures of numeracy (do not cause \FVR). In order
to reveal the possible scope of biases for plausible distributions
of \SWB, a number of parameters of the synthetic data generation process
were varied systematically. These include $\chi,$ $\boldsymbol{\boldsymbol{\beta}}_{N},$
 and two parameters determining the \emph{scale} and \emph{offset}
of the cut points.\footnote{See \href{https://alum.mit.edu/www/cpbl/publications/Barrington-Leigh-JPubE2024-appendix.pdf}{Appendix C} for details of the parameters used in the synthetic
data generating function, \href{https://alum.mit.edu/www/cpbl/publications/Barrington-Leigh-JPubE2024-appendix.pdf}{Appendix D} for some
detail from estimates using the simulated data, showing an example
of the complicated dependence of biases on attributes of the distribution,
and \href{https://alum.mit.edu/www/cpbl/publications/Barrington-Leigh-JPubE2024-appendix.pdf}{Appendix E} for propositions on the maximum
possible scope of these biases.} 

\begin{figure*}
\includegraphics[width=1\textwidth]{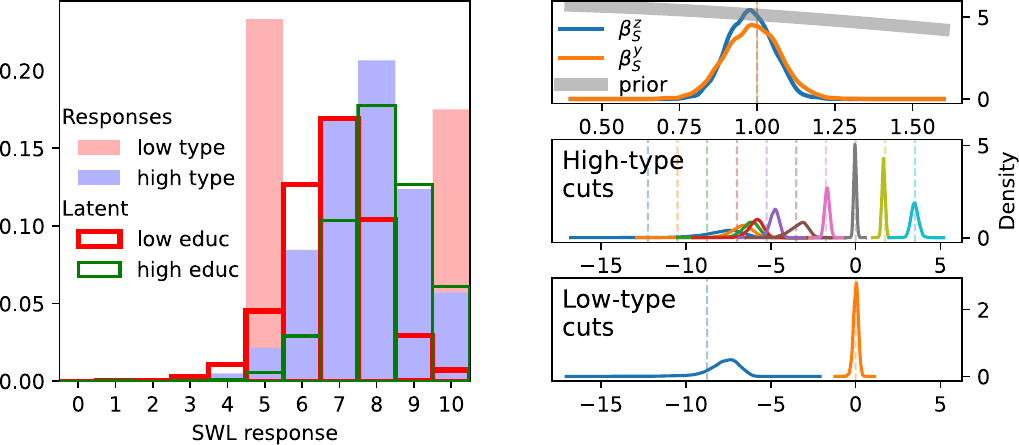}\\
\begin{tabular*}{1\textwidth}{@{\extracolsep{\fill}}>{\centering}p{0.333\textwidth}>{\centering}p{0.333\textwidth}}
(a) & (b)\tabularnewline
\end{tabular*}\caption[Example of synthetic data and validation]{Example of synthetic data and validation. Panel (a) shows simulated
and latent responses for one set of synthetic parameters. Overall
in this example, 33\% of respondents are low type. They reported an
average \SWB of 6.8, rather than their true average of 7.0, while high
types reported an average of 7.7. Panel (b) shows that the \FVR mixture
model correctly recovers synthetic coefficients $\boldsymbol{\boldsymbol{\beta}}_{S}$
and cut points $\boldsymbol{\boldsymbol{\alpha}}^{H}$, and $\boldsymbol{\boldsymbol{\alpha}}^{L}$.
Vertical dashed lines show the true (data generating process) values.
Most cut points are precisely estimated, but the lowest ones are poorly
constrained because there are few low responses for this particular
set of synthetic parameters.\label{fig:v3figa-bc}}
\end{figure*}

\figref{v3figa-bc} shows one example of a simulated distribution
of \SWB. In (a), shaded bars represent simulated responses on a 0 to
10 scale. Unlike in real data, we are able to identify which respondents
(among those giving a 0, 5, or 10) used \FVR. This portion of responses,
labeled ``low type'', are shaded pink. Also because the data are
synthetic, we are able to construct the latent (``true'') wellbeing
levels and thus the counterfactual 0--10 responses which would have
been given if everyone reported without \FVR. This counterfactual distribution,
including both low and high types, is shown split into two groups
based on education level. Although the true wellbeing distribution
of this sample is centered around 7.5, equidistant from the focal
values of 5 and 10, there is a net negative bias of $-0.08$ in mean
reported \SWB. This is because the distribution of the lower educated
component is generally closer to ``5'' than to ``10''. Thus, the
amount of rounding up is less than the amount of rounding down. 

Simulated biases in regression coefficients are obtained by estimating
a traditional ordered logit model on the synthetic data, and comparing
those estimates to the true values used in constructing the data,
$\beta_{S}^{z}=\beta_{S}^{y}=1$. In the case shown in \figref{v3figa-bc},
these biases are also both negative, namely $-14\%$ and $-23\%$
respectively.\footnote{The origin of these negative biases is slightly more subtle; see
\href{https://alum.mit.edu/www/cpbl/publications/Barrington-Leigh-JPubE2024-appendix.pdf}{Appendix D} and \href{https://alum.mit.edu/www/cpbl/publications/Barrington-Leigh-JPubE2024-appendix.pdf}{Appendix E}
for details.} 

Simulations were carried out for a wide range of parameters, generating
cases with both positive and negative biases on mean \SWL and on $\beta_{S}^{z}$
much larger than in this example. Simulated biases on $\beta_{S}^{y}$,
by contrast, tend to be negative.\footnote{ See \href{https://alum.mit.edu/www/cpbl/publications/Barrington-Leigh-JPubE2024-appendix.pdf}{Appendix D} for an explanation. In the
simulations, $y$ has no extra effect on (information about) numeracy,
after taking $z$ into account. In real data, income is likely to
contain variance that is informative for \FVR but orthogonal to available
measures of education. In this case, income in empirical applications
will exhibit a blend of the bias features attributed to $z$ and $y$
in these simulations.} More generally, the bias on mean \SWL can be as large as $\pm2$ points
(see \href{https://alum.mit.edu/www/cpbl/publications/Barrington-Leigh-JPubE2024-appendix.pdf}{Appendix Proposition E.1} in \href{https://alum.mit.edu/www/cpbl/publications/Barrington-Leigh-JPubE2024-appendix.pdf}{Appendix E})
and the bias on $\beta_{S}^{z}$ may be even larger (\href{https://alum.mit.edu/www/cpbl/publications/Barrington-Leigh-JPubE2024-appendix.pdf}{Appendix Proposition E.1}).
In all cases, the true distribution, fraction of low-types, and effects
of $z$ and $y$ on latent wellbeing are identified and correctly
estimated by the \FVR mixture model. As an example, \figref{v3figa-bc}(b)
shows estimated coefficients and cut points for the same case shown
in (a).

\subsection{Variance of SWL (``happiness inequality'')}

Although not a focus of this paper, it is also worth noting that variance
of \SWB, which has attracted interest as a measure of inequality \citep{Goff-Helliwell-Mayraz-EI2018-SWB-inequality,Hasegawa-Ueda-JSE2011-SWB-inequality,Stevenson-Wolfers-NBER2008-happiness-inequality-United-States},
also suffers from bias due to \FVR, as of course do other moments and
other measures of dispersion. For a relatively narrow distribution
of \SWL centred around 5, \FVR behavior decreases the variance. For
a wider distribution, focal values of 0 and 10 would become prominent,
and the variance could be biased upwards instead.

\section[Empirical estimates]{Empirical estimates of \FVR bias and \FVRI\label{sec:empirical-estimates}}

With the above evidence of parameter identification from simulated
data, the rest of this paper turns to empirical estimates. The distributions
of \SWL for different levels of education, shown in \figref{CCHS-17-18:SWL-by-education},
indicate the significance of focal value rounding behavior in the
CCHS sample. Using the mixture model, the role of education in supporting
\SWL can be estimated, despite the strong relationship between education
and the focal value bias. Columns (1) and (2) of \tabref{CCHSmixtureEstimates}
show the results of conventional, or ``naive'' estimates of the
following simple individual-level cross-sectional OLS model for \SWL,
 \iftwocolumncpbl
   \begin{align}
\text{SWL}_{i}=&c+\sum_{j}\beta_{S}^{h_{j}}\text{education}_{j,i}\nonumber \\ &+\beta_{S}^{I}\log\left(\text{HH income}\right)_{i}+\varepsilon_{i}
   \end{align}
\else
\[
\text{SWL}_{i}=c+\sum_{j}\beta_{S}^{h_{j}}\text{education}_{j,i}+\beta_{S}^{I}\log\left(\text{HH income}\right)_{i}+\varepsilon_{i}
\]
\fi
as well as its ordered logit counterpart. Educational attainment is
captured by a set of cumulative dummies, so that $\beta^{h_{j}}$
is the impact of having completed education level $j$ or higher.

The naive estimated coefficient on completing secondary education
is near-zero or distinctly negative in the two estimates. The ordered
logit coefficients predict that completion of high school \emph{reduces}
the odds of a higher \SWL by more than 7\%, and that even a university
education reduces those odds by nearly 4\% as compared with someone
who has less than a high school education. These values are economically
large; using the simultaneously-estimated coefficient on log income,
the former effect is estimated to be equivalent to a 13\% reduction
in income.\footnote{The values in this paragraph are calculated as $e^{-.075}-1=-0.072\approx-7\%$;
$e^{-.075+.038}-1=-0.036\approx-4\%$; and $e^{-.075/0.53}-1=-0.13\approx-13\%.$}

\begin{table*}
\ifusetabularsource  
  
    \renewcommand{\ctNtabCols}{5}
    \renewcommand{\ctFirstHeader}{\hline
 & \multicolumn{2}{c|}{Conventional} & \multicolumn{2}{c|}{Mixture model}\\ 

 & OLS & ologit & FVRI$\rightarrow$0 & Mixture\\ 

 & (1) & (2) & (3) & (4)\\ 
 \hline\hline
 }
    \renewcommand{\ctSubsequentHeaders}{\ctFirstHeader}
    \renewcommand{\ctBody}{Life satisfaction ($\beta_S$)&&&&\\  
&&&&\\  
\multicolumn{1}{|r|}{~ School: $\geq$Secondary}&$-$.002&\wrapSigOnePercent{$-$.075}&\wrapSigOneThousandth{$-$.073}&\wrapSigOnePercent{.060}\\  
&\coefse{.019}&\coefse{.023}&\coefse{.021}&\coefse{.024}\\  
\multicolumn{1}{|r|}{~ School: Post-secondary}&\wrapSigOneThousandth{.069}&\wrapSigFivePercent{.038}&\wrapSigOnePercent{.038}&\wrapSigOneThousandth{.17}\\  
&\coefse{.014}&\coefse{.015}&\coefse{.015}&\coefse{.018}\\  
\multicolumn{1}{|r|}{~ log(HH income)}&\wrapSigOneThousandth{.55}&\wrapSigOneThousandth{.53}&\wrapSigOneThousandth{.53}&\wrapSigOneThousandth{.62}\\  
&\coefse{.009}&\coefse{.011}&\coefse{.010}&\coefse{.011}\\  
\multicolumn{1}{|r|}{~ constant}&\wrapSigOneThousandth{2.0}&&&\\  
&\coefse{.093}&&&\\  
\hline Numeracy ($\beta_N$)&&&&\\  
&&&&\\  
\multicolumn{1}{|r|}{ constant}&&&&\wrapSigOneThousandth{1.09}\\  
&&&&\coefse{.043}\\  
\multicolumn{1}{|r|}{ School: $\geq$Secondary}&&&&\wrapSigOneThousandth{.45}\\  
&&&&\coefse{.038}\\  
\multicolumn{1}{|r|}{ School: Post-secondary}&&&&\wrapSigOneThousandth{.53}\\  
&&&&\coefse{.045}\\  
\multicolumn{1}{|r|}{ log(HH income)}&&&&\wrapSigOneThousandth{.22}\\  
&&&&\coefse{.028}\\  
\hline FVRI&&&&\wrapSigOneThousandth{.14}\\  
&&&&\coefse{.008}\\  
\hline&&&&\\  
obs.&91796&91796&91796&91796\\  
log likelihood&$-$177641&$-$160572&$-$160597&$-$159586\\ 
\cline{1-\ctNtabCols}
 }
    
    \renewcommand{\ctCaption}{ {\color{blue} minimal.... {\footnotesize\cpblColourLegend} } }
    \ifx\@ctUsingWrapper\@empty
    \begin{table}
    \begin{tabular}{|>{\bfseries}l|c|c|c|c|}
    \ctFirstHeader
    \ctBody
    \end{tabular}
    \end{table}
    \else
    \fi

    \renewcommand{\ctStartTabular}{\begin{tabular}{|>{\bfseries}l|c|c|c|c|}}
    \renewcommand{\ctStartLongtable}{\begin{longtable}[c]{|>{\bfseries}l|c|c|c|c|}}

 \begin{minipage}{0.8\textwidth}    
            \ctStartTabular
            \ctFirstHeader
            \ctBody \hline
          \ctEndTabular%
          \end{minipage}~\begin{minipage}{0.4\textwidth}    
       \includegraphics[height=13cm]{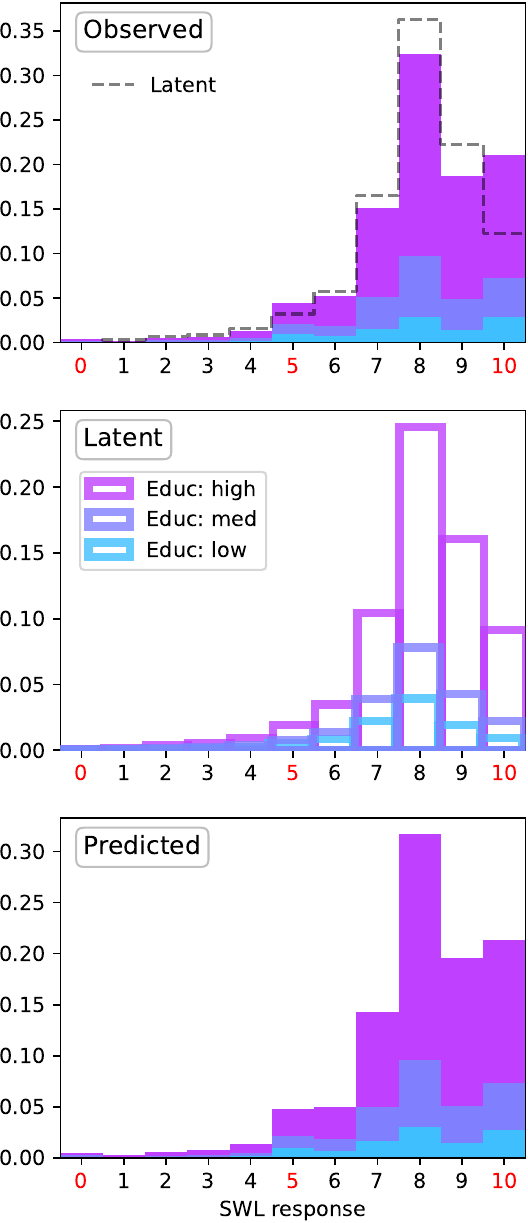}
                      \end{minipage} 
                      {\footnotesize\cpblColourLegend}
                    \else
    \includegraphics[height=11cm]{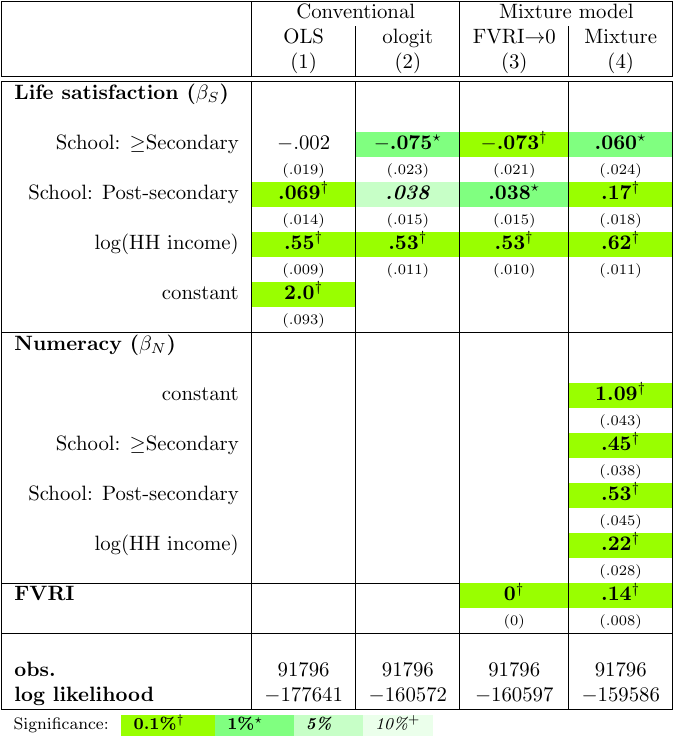}\enspace{}\includegraphics[height=11cm]{jpube-assets//postestimateHists-CCHS1718-MODEL21-7867763196}
\fi                      
\centering{}\caption[Estimates of life satisfaction in CCHS]{Estimates of life satisfaction in CCHS. The first two columns show
conventional, \textquotedblleft naive\textquotedblright{} estimates
(as raw coefficients) of a model explaining life satisfaction with
just education and household income. Column (3) shows estimates from
a degenerate version of the mixture model constrained to exclude \FVR
behavior. Column (4) shows the unconstrained mixture model estimate,
with significantly higher effects of education and income on life
satisfaction. Histograms show response distributions split up (and
colored) by education. The top plot is observed values, with the model's
inferred overall latent distribution shown by a dashed line. The second
and third show latent (or \textquotedblleft corrected\textquotedblright )
and predicted responses.\label{tab:CCHSmixtureEstimates}}
\end{table*}

When constrained to disallow focal value behavior, the mixture model's
estimate, shown in column (3) of \tabref{CCHSmixtureEstimates},
reproduces the ordered logit values, as it should. However, when the
full model is estimated, a significantly positive value ($\sim$0.06)
is found for the \SWL benefit of completion of secondary school, and
an additional 0.17 for those completing post-secondary.

The bias in a conventional estimate of the income coefficient is also
large: the mixture model strongly rejects the naive estimated value
of $\sim$0.53, in favor of a value of $\sim$0.62. This represents
a 17\% difference in the most studied value in happiness economics.
 Combining these coefficients implies that, after controlling for
income, the true benefit of college completion, as compared with an
otherwise-similar respondent without high school completion, is equivalent
to an additional 45\% of income. High school completion by itself
confers a benefit equivalent to more than 10\% of income, after controlling
for differences in actual income.

The specification in \tabref{CCHSmixtureEstimates} includes both
education and income as predictors of \FVR. \href{https://alum.mit.edu/www/cpbl/publications/Barrington-Leigh-JPubE2024-appendix.pdf}{Appendix Table F.2}
shows that alternate models with only education in the \FVR equation,
or with additional controls, give highly consistent results.

Next to \tabref{CCHSmixtureEstimates} are visualizations of several
sets of distributions, showing the model's ability to predict observed
response patterns while estimating the distribution of ``underlying''
or ``true'' \SWB.

\section{Applications\label{sec:Applications}}

Hundreds of empirical papers estimating models of life satisfaction
and other extended-Likert-like scales could be revisited in light
of the significant possibility of biases identified above. Those focusing
on effects of socioeconomic status, gender, and age, and those which
particularly address populations with low levels of numeracy, especially
invite reanalysis. Here I reproduce estimates from three papers to
exemplify the important changes that may result from such analysis,
and to show that the often-reproduced ``paradox'' of negative benefits
to education may be largely resolved by the cognitive mixture model.

\subsection{U.K.: Clark and Oswald (1996)\label{sec:Clark-and-Oswald}}

The first of these papers, with over 1500 citations, is a relatively
early contribution in the modern study of relative income concerns
but also prominently points out the anomalously low estimated returns
to wellbeing from education \citep{Clark-Oswald-JPubE1996}. It was
also recently cited as one of 11 studies in the major accumulated
evidence on the life satisfaction benefits from additional education
\citep[see Annex 3a]{Clark-Fleche-Layard-Powdthavee-Ward-2019-origins-of-happiness}.
In fact, the paper uses data from the British Household Panel Survey
(BHPS) prior to its inclusion of \SWL, so it uses instead responses
to 7-point satisfaction with pay and satisfaction with job questions.

\begin{figure*}
\begin{centering}
\includegraphics[width=1\textwidth]{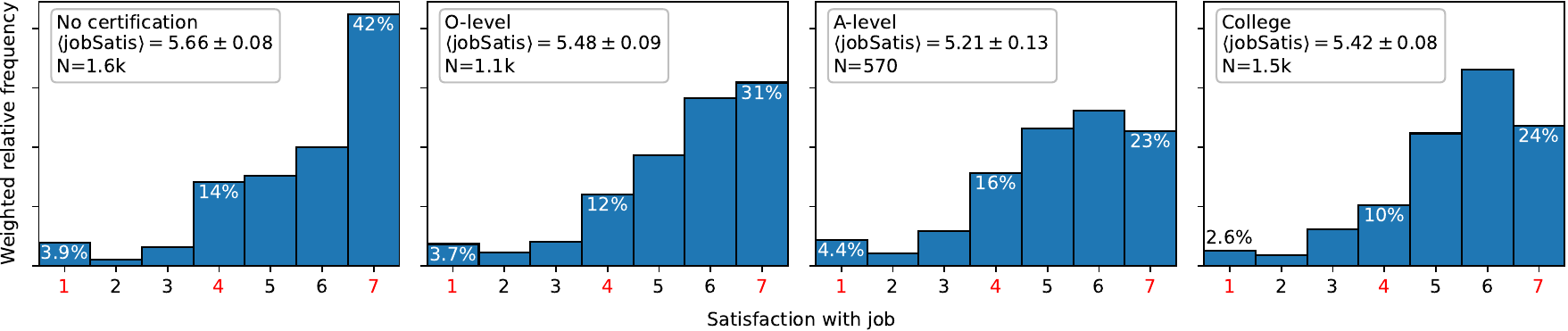}
\par\end{centering}
~\\

\begin{centering}
\includegraphics[width=1\textwidth]{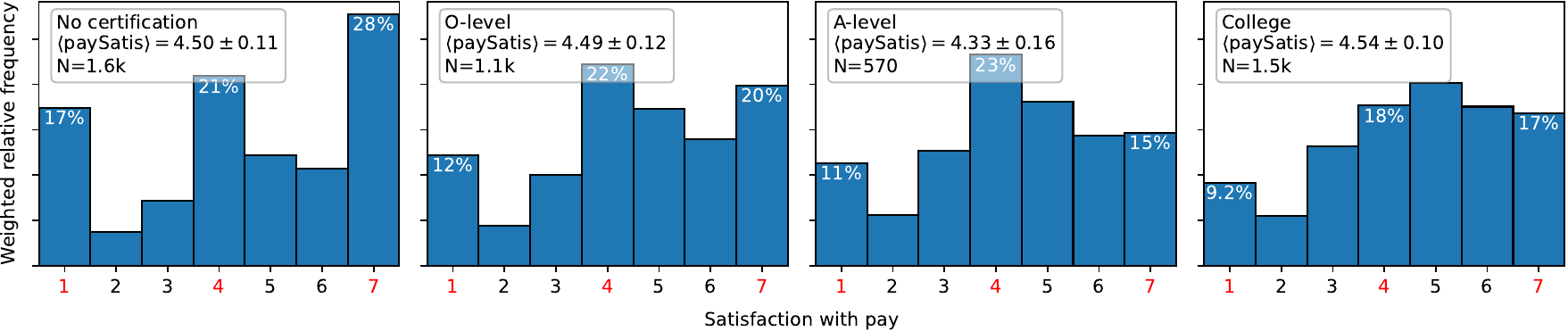}
\par\end{centering}
\centering{}\caption[Distributions by education level of satisfaction with job and with
pay]{Distributions of satisfaction with job and with pay for different
categories of educational attainment.\label{fig:BHPS-SWB-by-education}}
\end{figure*}

\citet{Clark-Oswald-JPubE1996} did not examine the distributions
of these subjective response variables according to formal educational
attainment.\footnote{The description from \citet{Clark-Oswald-JPubE1996} reads: ``Table
5 contains two ordered probits, in each of which three dummies for
educational attainment are included as well as a control for income.
The dummies are for a college degree, advanced high school (A-level
approximately), and intermediate high school (O-level approximately).
The omitted category is for no or low qualifications. These four categories
are for achieved paper certificates and not merely for years of schooling''.} Doing so reveals dramatic \FVR behavior which roughly diminishes with
education (\figref{BHPS-SWB-by-education}). The distribution of
satisfaction with pay is wider and more central (i.e., near ``4'')
than that of job satisfaction, and features unmistakable evidence
of all three focal values (1, 4, and 7) for groups with lower academic
certifications. For job satisfaction, the upper focal value is most
obvious but all three are evident on inspection. If those with A-levels
but no College are excluded, then the group means and the prevalence
of each focal value all decrease monotonically with education. 

\tabref{jobSatis-BHPS-main-estimates} shows raw coefficients for
model estimates of overall satisfaction with job. The first three
models are conventional estimation approaches, including an ordered
probit model, which nearly reproduces the published values\footnote{The coefficient shown on log income (.016) strongly disagrees with
the published value (.50) in \citet{Clark-Oswald-JPubE1996}. Upon
contacting the authors, it was determined that a typo in production
of the original work resulted in a reporting of 0.50 rather than the
estimated 0.05 for this coefficient (personal communication, Andrew
Clark, 2021). This error has not been previously reported. Because
of the error, the authors did not address the surprisingly low coefficient
on log of household income. The set of regional, health, race, industry,
and occupation dummies are excluded in \tabref{paySatis-BHPS-main-estimates}
because the exact definitions from the 1996 work are not available.} and retained sample size (4730 in all my estimates) of the main estimate
in \citet[Table 5]{Clark-Oswald-JPubE1996}.\footnote{For easier comparison with their table, the education categories are
mutually exclusive, rather than cumulative, as in the CCHS data and
the data to follow.} In ordered probit, OLS, and ordered logit models, academic attainment
is strongly predictive of \emph{lower} satisfaction after adjusting
for log of household income. The implied effect is enormous. As compared
with someone with primary education only, an advanced high school
graduate (A-levels) is less satisfied with their job by as much as
they would be with a 3-fold \emph{decrease} in wage.\footnote{The coefficients on log job hours and on A-levels education are nearly
identical, implying that, having already adjusted for income, a unit
log, or factor $\sim$2.7, increase in hours worked predicts a similar
change to job satisfaction as does the educational attainment.} As already shown in \figref{BHPS-SWB-by-education}, even the raw
mean job satisfaction is decreasing across the first three education
groups. \citet{Clark-Oswald-JPubE1996} speculate that their findings
of low satisfaction of the higher educated may be related to a recent
recession that particularly hit the middle class in the UK, but also
cite several earlier studies which corroborate the negative or negligible
benefits from education on job satisfaction.

\begin{table*}
  \begin{centering}
    \ifusetabularsource
      
    \renewcommand{\ctNtabCols}{6}
    \renewcommand{\ctFirstHeader}{\hline
 & \multicolumn{3}{c|}{Conventional} & \multicolumn{2}{c|}{Mixture model}\\ 

 & oprobit & OLS & ologit & FVRI$\rightarrow$0 & Mixture\\ 

 & (1) & (2) & (3) & (4) & (5)\\ 
 \hline\hline
 }
    \renewcommand{\ctSubsequentHeaders}{\ctFirstHeader}
    \renewcommand{\ctBody}{Job Satisfaction ($\beta_S$)&&&&&\\  
&&&&&\\  
\multicolumn{1}{|r|}{~ log(HH income)}&.016&.074&.015&.006&\wrapSigFivePercent{.14}\\  
&\coefse{.036}&\coefse{.049}&\coefse{.060}&\coefse{.060}&\coefse{.076}\\  
\multicolumn{1}{|r|}{~ {\sc educ:} College degree}&\wrapSigOneThousandth{$-$.15}&\wrapSigFivePercent{$-$.14}&\wrapSigOneThousandth{$-$.27}&\wrapSigOneThousandth{$-$.25}&.001\\  
&\coefse{.046}&\coefse{.061}&\coefse{.078}&\coefse{.075}&\coefse{.11}\\  
\multicolumn{1}{|r|}{~ {\sc educ:} A-levels (approx)}&\wrapSigOneThousandth{$-$.26}&\wrapSigOneThousandth{$-$.31}&\wrapSigOneThousandth{$-$.45}&\wrapSigOneThousandth{$-$.43}&\wrapSigFivePercent{$-$.24}\\  
&\coefse{.055}&\coefse{.076}&\coefse{.095}&\coefse{.094}&\coefse{.11}\\  
\multicolumn{1}{|r|}{~ {\sc educ:} O-levels (approx)}&\wrapSigFivePercent{$-$.10}&$-$.096&\wrapSigFivePercent{$-$.19}&\wrapSigOnePercent{$-$.18}&.013\\  
&\coefse{.046}&\coefse{.060}&\coefse{.078}&\coefse{.075}&\coefse{.092}\\  
\multicolumn{1}{|r|}{~ Log job hours}&\wrapSigOneThousandth{$-$.27}&\wrapSigOneThousandth{$-$.35}&\wrapSigOneThousandth{$-$.45}&\wrapSigOneThousandth{$-$.44}&\wrapSigOneThousandth{$-$.46}\\  
&\coefse{.054}&\coefse{.071}&\coefse{.092}&\coefse{.088}&\coefse{.099}\\  
\multicolumn{1}{|r|}{~ age}&\wrapSigOneThousandth{$-$.033}&\wrapSigOnePercent{$-$.037}&\wrapSigOneThousandth{$-$.055}&\wrapSigOneThousandth{$-$.051}&\wrapSigOnePercent{$-$.051}\\  
&\coefse{.009}&\coefse{.012}&\coefse{.015}&\coefse{.015}&\coefse{.017}\\  
\multicolumn{1}{|r|}{~ age$^2$/1000}&\wrapSigOneThousandth{.53}&\wrapSigOneThousandth{.59}&\wrapSigOneThousandth{.88}&\wrapSigOneThousandth{.83}&\wrapSigOneThousandth{.84}\\  
&\coefse{.11}&\coefse{.15}&\coefse{.19}&\coefse{.18}&\coefse{.21}\\  
\multicolumn{1}{|r|}{~ female}&\wrapSigOneThousandth{.24}&\wrapSigOneThousandth{.35}&\wrapSigOneThousandth{.41}&\wrapSigOneThousandth{.41}&\wrapSigOneThousandth{.45}\\  
&\coefse{.036}&\coefse{.050}&\coefse{.061}&\coefse{.062}&\coefse{.069}\\  
\multicolumn{1}{|r|}{~ constant}&&\wrapSigOneThousandth{6.3}&&&\\  
&&\coefse{.32}&&&\\  
\hline Numeracy ($\beta_N$)&&&&&\\  
&&&&&\\  
\multicolumn{1}{|r|}{ constant}&&&&&\wrapSigOnePercent{.41}\\  
&&&&&\coefse{.16}\\  
\multicolumn{1}{|r|}{ log(HH income)}&&&&&\wrapSigOneThousandth{.49}\\  
&&&&&\coefse{.084}\\  
\multicolumn{1}{|r|}{ {\sc educ:} College degree}&&&&&\wrapSigOneThousandth{1.30}\\  
&&&&&\coefse{.28}\\  
\multicolumn{1}{|r|}{ {\sc educ:} A-levels (approx)}&&&&&\wrapSigOneThousandth{.86}\\  
&&&&&\coefse{.21}\\  
\multicolumn{1}{|r|}{ {\sc educ:} O-levels (approx)}&&&&&\wrapSigOneThousandth{.69}\\  
&&&&&\coefse{.14}\\  
\hline FVRI&&&&\wrapSigOneThousandth{0}&\wrapSigOneThousandth{.28}\\  
&&&&\coefse{0}&\coefse{.036}\\  
\hline&&&&&\\  
obs.&4730&4730&4730&4730&4730\\  
log likelihood&$-$7522&$-$8573&$-$7516&$-$7531&$-$7466\\ 
\cline{1-\ctNtabCols}
 }
    
    \renewcommand{\ctCaption}{ {\color{blue} minimal.... {\footnotesize\cpblColourLegend} } }
    \ifx\@ctUsingWrapper\@empty
    \begin{table}
    \begin{tabular}{|>{\bfseries}l|c|c|c|c|c|}
    \ctFirstHeader
    \ctBody
    \end{tabular}
    \end{table}
    \else
    \fi

    \renewcommand{\ctStartTabular}{\begin{tabular}{|>{\bfseries}l|c|c|c|c|c|}}
    \renewcommand{\ctStartLongtable}{\begin{longtable}[c]{|>{\bfseries}l|c|c|c|c|c|}}

            \ctStartTabular
            \ctFirstHeader
            \ctBody \hline
          \ctEndTabular%
      \else
        \includegraphics[height=14cm]{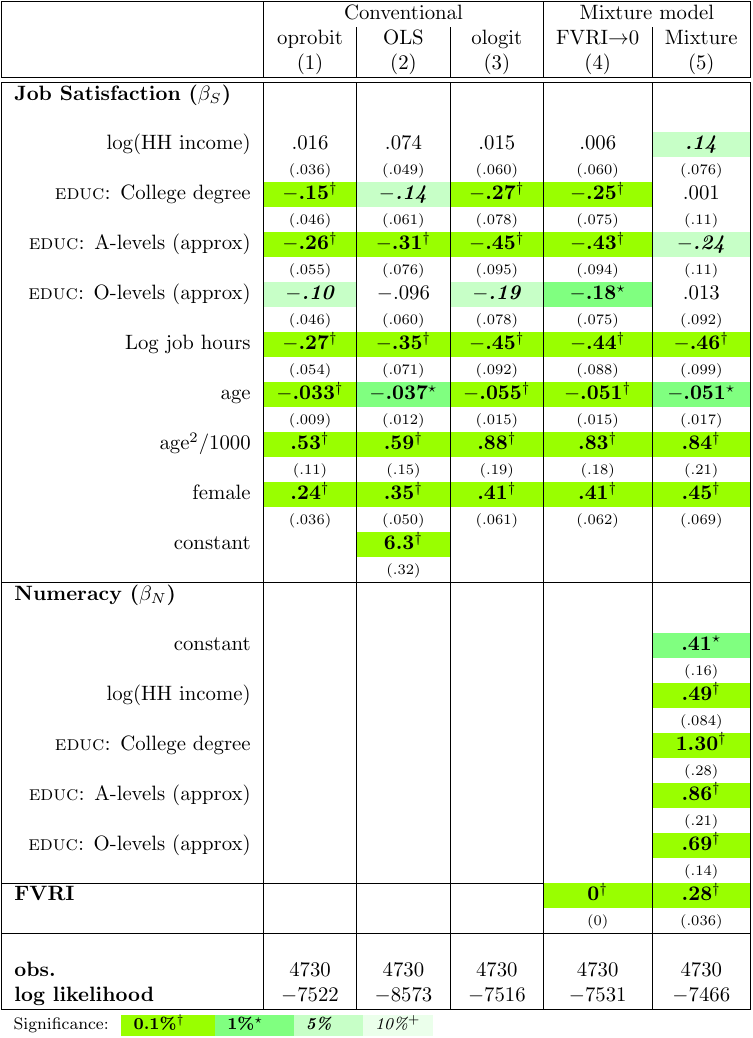}\par
      \fi
\end{centering}
\caption[Estimates of job satisfaction in BHPS]{Estimates of job satisfaction in BHPS. Raw coefficients are shown.
Education indicators identify mutually exclusive groups in comparison
to those with less than O-levels.\label{tab:jobSatis-BHPS-main-estimates}}
\end{table*}

Equally surprising in these results is the nil effect of income on
job satisfaction. The 95\% confidence interval for the coefficient
of log income in column (3) is  $-$0.10 to $+$0.13, with the upper
limit implying that a \emph{doubling} of income would increase the
odds of a higher satisfaction response by less than 10\%.

\begin{table*}
  \begin{centering}
    \ifusetabularsource
      
    \renewcommand{\ctNtabCols}{6}
    \renewcommand{\ctFirstHeader}{\hline
 & \multicolumn{3}{c|}{Conventional} & \multicolumn{2}{c|}{Mixture model}\\ 

 & oprobit & OLS & ologit & FVRI$\rightarrow$0 & Mixture\\ 

 & (1) & (2) & (3) & (4) & (5)\\ 
 \hline\hline
 }
    \renewcommand{\ctSubsequentHeaders}{\ctFirstHeader}
    \renewcommand{\ctBody}{Pay Satisfaction ($\beta_S$)&&&&&\\  
&&&&&\\  
\multicolumn{1}{|r|}{~ log(HH income)}&\wrapSigOneThousandth{.50}&\wrapSigOneThousandth{.92}&\wrapSigOneThousandth{.88}&\wrapSigOneThousandth{.87}&\wrapSigOneThousandth{.96}\\  
&\coefse{.038}&\coefse{.062}&\coefse{.065}&\coefse{.059}&\coefse{.068}\\  
\multicolumn{1}{|r|}{~ {\sc educ:} College degree}&\wrapSigOneThousandth{$-$.17}&\wrapSigOneThousandth{$-$.26}&\wrapSigOneThousandth{$-$.31}&\wrapSigOneThousandth{$-$.30}&\wrapSigOnePercent{$-$.26}\\  
&\coefse{.045}&\coefse{.077}&\coefse{.078}&\coefse{.071}&\coefse{.089}\\  
\multicolumn{1}{|r|}{~ {\sc educ:} A-levels (approx)}&\wrapSigOnePercent{$-$.14}&\wrapSigFivePercent{$-$.20}&\wrapSigOnePercent{$-$.25}&\wrapSigOnePercent{$-$.24}&\wrapSigFivePercent{$-$.20}\\  
&\coefse{.053}&\coefse{.097}&\coefse{.090}&\coefse{.091}&\coefse{.10}\\  
\multicolumn{1}{|r|}{~ {\sc educ:} O-levels (approx)}&$-$.029&$-$.019&$-$.051&$-$.042&.005\\  
&\coefse{.045}&\coefse{.077}&\coefse{.077}&\coefse{.072}&\coefse{.081}\\  
\multicolumn{1}{|r|}{~ Log job hours}&\wrapSigOneThousandth{$-$.82}&\wrapSigOneThousandth{$-$1.42}&\wrapSigOneThousandth{$-$1.44}&\wrapSigOneThousandth{$-$1.43}&\wrapSigOneThousandth{$-$1.46}\\  
&\coefse{.058}&\coefse{.089}&\coefse{.10}&\coefse{.091}&\coefse{.094}\\  
\multicolumn{1}{|r|}{~ age}&\wrapSigOneThousandth{$-$.043}&\wrapSigOneThousandth{$-$.072}&\wrapSigOneThousandth{$-$.077}&\wrapSigOneThousandth{$-$.071}&\wrapSigOneThousandth{$-$.074}\\  
&\coefse{.009}&\coefse{.015}&\coefse{.015}&\coefse{.014}&\coefse{.015}\\  
\multicolumn{1}{|r|}{~ age$^2$/1000}&\wrapSigOneThousandth{.62}&\wrapSigOneThousandth{1.03}&\wrapSigOneThousandth{1.10}&\wrapSigOneThousandth{1.03}&\wrapSigOneThousandth{1.06}\\  
&\coefse{.11}&\coefse{.19}&\coefse{.19}&\coefse{.18}&\coefse{.18}\\  
\multicolumn{1}{|r|}{~ female}&\wrapSigOneThousandth{.27}&\wrapSigOneThousandth{.48}&\wrapSigOneThousandth{.45}&\wrapSigOneThousandth{.44}&\wrapSigOneThousandth{.45}\\  
&\coefse{.035}&\coefse{.064}&\coefse{.059}&\coefse{.059}&\coefse{.064}\\  
\multicolumn{1}{|r|}{~ constant}&&\wrapSigOneThousandth{3.8}&&&\\  
&&\coefse{.40}&&&\\  
\hline Numeracy ($\beta_N$)&&&&&\\  
&&&&&\\  
\multicolumn{1}{|r|}{ constant}&&&&&\wrapSigTenPercent{.22}\\  
&&&&&\coefse{.17}\\  
\multicolumn{1}{|r|}{ log(HH income)}&&&&&\wrapSigOneThousandth{.49}\\  
&&&&&\coefse{.075}\\  
\multicolumn{1}{|r|}{ {\sc educ:} College degree}&&&&&\wrapSigOneThousandth{1.26}\\  
&&&&&\coefse{.23}\\  
\multicolumn{1}{|r|}{ {\sc educ:} A-levels (approx)}&&&&&\wrapSigOneThousandth{1.10}\\  
&&&&&\coefse{.24}\\  
\multicolumn{1}{|r|}{ {\sc educ:} O-levels (approx)}&&&&&\wrapSigOneThousandth{.72}\\  
&&&&&\coefse{.15}\\  
\hline FVRI&&&&\wrapSigOneThousandth{0}&\wrapSigOneThousandth{.31}\\  
&&&&\coefse{0}&\coefse{.041}\\  
\hline&&&&&\\  
obs.&4730&4730&4730&4730&4730\\  
log likelihood&$-$8642&$-$9689&$-$8635&$-$8648&$-$8540\\ 
\cline{1-\ctNtabCols}
 }
    
    \renewcommand{\ctCaption}{ {\color{blue} minimal.... {\footnotesize\cpblColourLegend} } }
    \ifx\@ctUsingWrapper\@empty
    \begin{table}
    \begin{tabular}{|>{\bfseries}l|c|c|c|c|c|}
    \ctFirstHeader
    \ctBody
    \end{tabular}
    \end{table}
    \else
    \fi

    \renewcommand{\ctStartTabular}{\begin{tabular}{|>{\bfseries}l|c|c|c|c|c|}}
    \renewcommand{\ctStartLongtable}{\begin{longtable}[c]{|>{\bfseries}l|c|c|c|c|c|}}

            \ctStartTabular
            \ctFirstHeader
            \ctBody \hline
          \ctEndTabular%
      \else
\includegraphics[height=14cm]{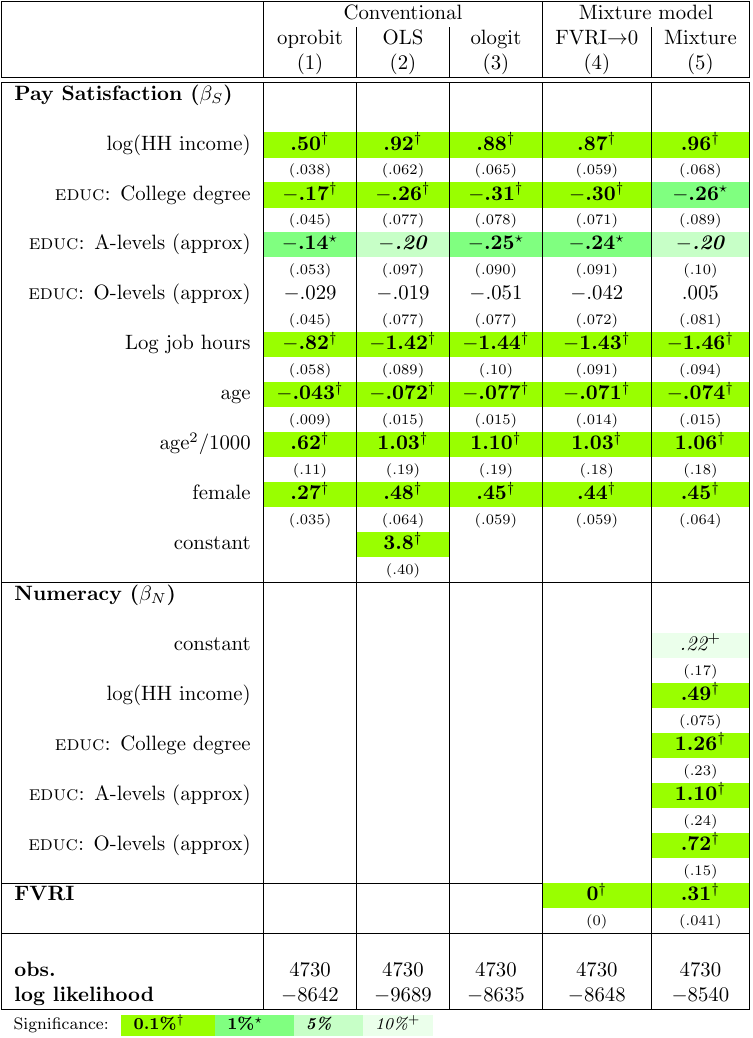}
\par\fi
\end{centering}
\caption[Estimates of satisfaction with pay in BHPS]{Estimates of satisfaction with pay in BHPS. Description as for \tabref{jobSatis-BHPS-main-estimates}.\label{tab:paySatis-BHPS-main-estimates}}
\end{table*}

Column (4) simply shows that the cognitive mixture model reproduces
an ordered logit estimate when focal value behavior is turned off,
while the key result lies in Column (5). When focal value behavior
is accounted for, the income coefficient increases to a confidently
positive value, and the strongly negative coefficients on O-level
and College completion are eliminated.  Respondents who finished
A-levels but stopped there for some reason, i.e., did not complete
college, are still predicted to be less satisfied with their jobs,
but the effect is half as large as in the naive model. Estimates of
other coefficients remain statistically unchanged. Both formal education
and reported income prove significant in predicting focal value behavior.
The estimated fraction of respondents, overall, who restricted their
answers to focal values is 28\%. The model also estimates a significant
bias in the mean reported job satisfaction, from a latent value of
5.3 which would have obtained had all respondents used the full scale,
to the observed value of 5.5. The model estimates that the low-numeracy
(\FVR) respondents reported an average job satisfaction of 5.9, and
that the high-numeracy respondents reported an average of 5.3.

\tabref{paySatis-BHPS-main-estimates} parallels \tabref{jobSatis-BHPS-main-estimates}
but relates to the other column in \citet{Clark-Oswald-JPubE1996}'s
Table 5 --- an estimate for satisfaction with\emph{ pay} rather than
with the \emph{job} overall. In this case, increased income is a strong
predictor of satisfaction even in naive estimates. On the other hand,
higher education again strongly predicts lower satisfaction, after
adjusting for household income, in conventional models. This may make
sense if the primary effect of education in this context is to set
expectations about pay. In any case, for satisfaction with pay, the
\FVR mixture model corroborates the estimates of the naive ordered
logit model.

How can the coefficient estimates remain relatively unchanged in the
presence of such a high degree of \FVR? While column (5) of \tabref{paySatis-BHPS-main-estimates}
shows that income and higher education levels predict lower propensity
for \FVR, and that 31\% of respondents used a simplified response scale
for answering this question, the net effect of \FVR on the estimated
coefficients is small. This can be understood by considering the distribution
of latent wellbeing values, with reference to the discussion in \secref{Validation}
and the Remark for \href{https://alum.mit.edu/www/cpbl/publications/Barrington-Leigh-JPubE2024-appendix.pdf}{Appendix Proposition E.2}. For this sample,
the number of respondents rounding up from 6 to 7 or from 3 to 4 is
balanced by the number rounding down from 2 to 1 or from 5 to 4.\footnote{As discussed earlier and as this example shows, there is no simple
relationship between the extent of \FVR and the size of net biases,
due to the possibility of offsetting contributions to bias and the
importance of detailed distributional features of the sample. It is
also worth noting that the model is capable of accounting for a high
fraction of extreme values (1s and 7s, here) as scale boundary effects
rather than \FVR. Indeed, it is also capable of accounting for a central
peak (here, at 4) without appealing to the existence of any \FVR. Instead,
the model estimate suggests that respondents were simplifying the
scale, and that the independently-estimated fractions of respondents
who did so were the same (28\% and 31\%) for the two questions.} \href{https://alum.mit.edu/www/cpbl/publications/Barrington-Leigh-JPubE2024-appendix.pdf}{Appendix Figure F.3} shows the estimated
distributions of responses which would have been given in the absence
of any \FVR (second row), for both job and pay satisfaction. All education
levels exhibit broad distributions of latent pay satisfaction, and
all carried out some degree of \FVR. 

\subsection{Australia: \citet{Powdthavee-Lekfuangfu-Wooden-JBEE2015-education-SWL-Australia-HILDA}}

More recently, \citet{Powdthavee-Lekfuangfu-Wooden-JBEE2015-education-SWL-Australia-HILDA}
have shed some further light on the apparent negative or insignificant
returns to education in life satisfaction regressions. They articulate
a more considered causal model for the impact of educational attainment
on overall life evaluations, taking into account several of the multiple
non-monetary channels through which education is expected or known
to affect life. In particular, they allow for mediating effects of
education through health, marriage, child-rearing, and employment,
in addition to income. They conclude that ``education is likely to
be positively related to overall life satisfaction through many different
channels, even when \emph{ceteris paribus} education itself has a
negative and statistically significant relationship with overall life
satisfaction''. Thus, while identifying some positive indirect effects
of education, their analysis does not account for the overall negative
effect of education on life satisfaction.

Here I do not integrate their panel data mediation pathways into the
\FVR model, which would go beyond the scope of this paper. Instead,
I use one cycle (2010) of the HILDA survey \citep[see][]{Powdthavee-Lekfuangfu-Wooden-JBEE2015-education-SWL-Australia-HILDA}
to test the same questions as above: how much of the negative overall
association between education and life satisfaction is accounted for
by \FVR behaviour? and how biased is the income coefficient when \FVR
is ignored?

\begin{figure*}
\includegraphics[width=1\textwidth]{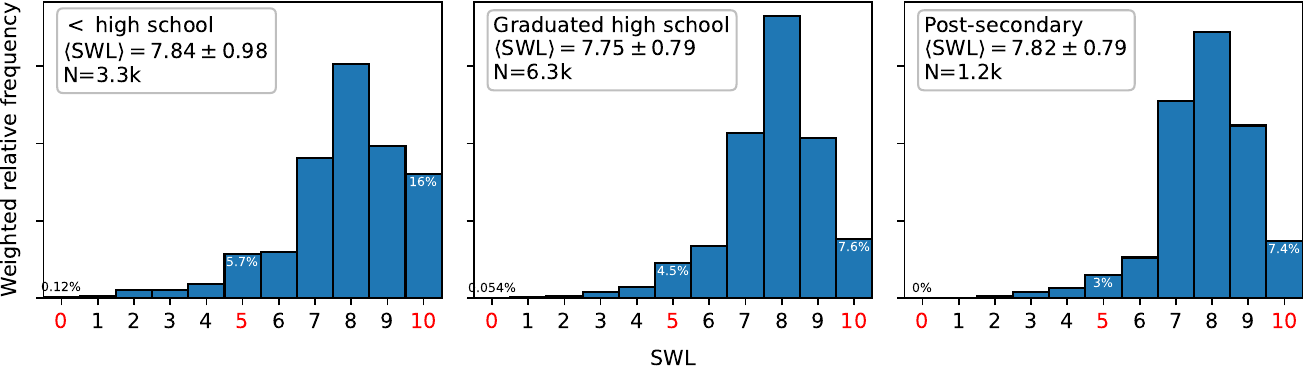}\caption[SWL responses in Australia]{\SWL responses in Australia by education level for those over 25 years
of age\label{fig:histograms-HILDA-by-education-25yrold}}
\end{figure*}

\figref{histograms-HILDA-by-education-25yrold} shows a familiar pattern
in weighted life satisfaction response distributions when separated
by education level. Here the focal value enhancements are more subtle,
but anomalously high response fractions for 5 and 10 are noticeable
at least in the lowest education group, and the proportions of each
focal value decrease across education groups. 

\tabref{HILDA-main-estimates} shows the comparison in now-familiar
form of the naive estimates of income and education effects on life
satisfaction in Australia (columns 1, 2, and 3) with an estimate of
the \FVR model in column (4). In the \FVR-aware model, the coefficient
on income approximately doubles, jumping by 4 standard errors. The
additional effect of college degree attainment after finishing high
school becomes weakly positive, and the effect of high school graduation
climbs by 5 standard errors.

\begin{table*}
  \centering
    \ifusetabularsource
      
    \renewcommand{\ctNtabCols}{5}
    \renewcommand{\ctFirstHeader}{\hline
 & \multicolumn{2}{c|}{Conventional} & \multicolumn{2}{c|}{Mixture model}\\ 

 & OLS & ologit & FVRI$\rightarrow$0 & Mixture\\ 

 & (1) & (2) & (3) & (4)\\ 
 \hline\hline
 }
    \renewcommand{\ctSubsequentHeaders}{\ctFirstHeader}
    \renewcommand{\ctBody}{Life satisfaction ($\beta_S$)&&&&\\  
&&&&\\  
\multicolumn{1}{|r|}{~ log(HH income)}&\wrapSigOneThousandth{.13}&\wrapSigOneThousandth{.11}&\wrapSigOneThousandth{.10}&\wrapSigOneThousandth{.21}\\  
&\coefse{.019}&\coefse{.027}&\coefse{.024}&\coefse{.026}\\  
\multicolumn{1}{|r|}{~ educHigh}&\wrapSigOneThousandth{$-$.15}&\wrapSigOneThousandth{$-$.27}&\wrapSigOneThousandth{$-$.26}&$-$.055\\  
&\coefse{.032}&\coefse{.044}&\coefse{.041}&\coefse{.045}\\  
\multicolumn{1}{|r|}{~ educCollege}&.042&.055&.054&\wrapSigTenPercent{.087}\\  
&\coefse{.047}&\coefse{.052}&\coefse{.057}&\coefse{.059}\\  
\multicolumn{1}{|r|}{~ constant}&\wrapSigOneThousandth{6.4}&&&\\  
&\coefse{.21}&&&\\  
\hline Numeracy ($\beta_N$)&&&&\\  
&&&&\\  
\multicolumn{1}{|r|}{ constant}&&&&\wrapSigOneThousandth{1.47}\\  
&&&&\coefse{.074}\\  
\multicolumn{1}{|r|}{ educHigh}&&&&\wrapSigOneThousandth{.93}\\  
&&&&\coefse{.081}\\  
\multicolumn{1}{|r|}{ educCollege}&&&&\wrapSigFivePercent{.30}\\  
&&&&\coefse{.15}\\  
\hline FVRI&&&\wrapSigOneThousandth{0}&\wrapSigOneThousandth{.11}\\  
&&&\coefse{0}&\coefse{.009}\\  
\hline&&&&\\  
obs.&10744&10744&10744&10744\\  
log likelihood&$-$19290&$-$18185&$-$18211&$-$18104\\ 
\cline{1-\ctNtabCols}
 }
    
    \renewcommand{\ctCaption}{ {\color{blue} minimal.... {\footnotesize\cpblColourLegend} } }
    \ifx\@ctUsingWrapper\@empty
    \begin{table}
    \begin{tabular}{|>{\bfseries}l|c|c|c|c|}
    \ctFirstHeader
    \ctBody
    \end{tabular}
    \end{table}
    \else
    \fi

    \renewcommand{\ctStartTabular}{\begin{tabular}{|>{\bfseries}l|c|c|c|c|}}
    \renewcommand{\ctStartLongtable}{\begin{longtable}[c]{|>{\bfseries}l|c|c|c|c|}}

            \ctStartTabular
            \ctFirstHeader
            \ctBody \hline
          \ctEndTabular%
      \else
        \includegraphics[height=9cm]{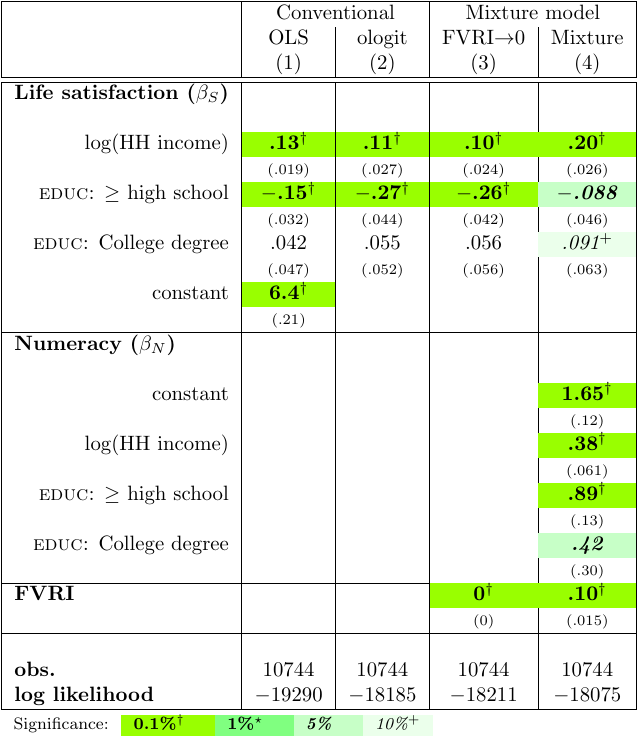}
      \fi
      \caption[Estimates of life satisfaction in HILDA]{Estimates of life satisfaction in the 2010 cycle of HILDA. \label{tab:HILDA-main-estimates}}
\end{table*}

\subsection{First Nations and Métis in Canada\label{sec:indigenous}}

Next I pick on my own prior work by re-examining a paper which reported
an anomalously low benefit of income for a sample of  Indigenous
(First Nations and Métis) peoples in Canada \citep{Barrington-Leigh-Sloman-IIPJ2016-aboriginal}.
In addition to estimating a negative effect of income on life satisfaction,
we found an average life satisfaction among Indigenous respondents
that was equivalent to that of the general population, despite the
stark objective challenges faced by the former groups, including disproportionate
levels of discrimination and socioeconomic disadvantage with respect
to the rest of the Canadian population. \citet{Barrington-Leigh-Sloman-IIPJ2016-aboriginal}
suggested as a possible interpretation that total income is not well
measured by the standard income question for this group, but remain
``cautious and skeptical'' about the data overall.

\begin{figure*}
\begin{centering}
\includegraphics[width=0.7\textwidth]{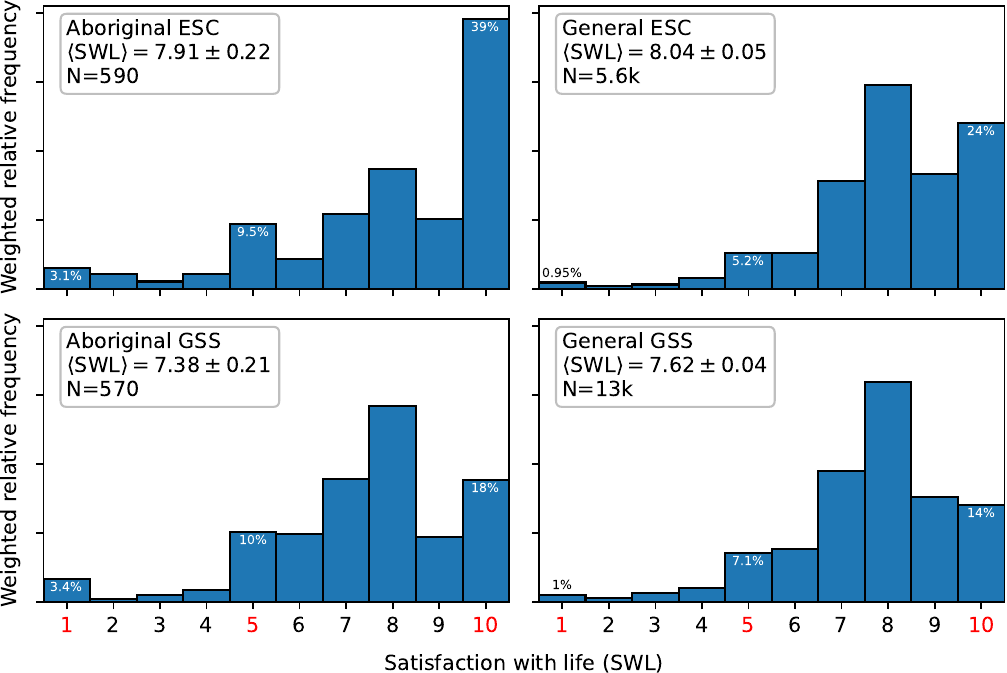}
\par\end{centering}
\caption[Life satisfaction of Indigenous Canadians]{Life satisfaction of Indigenous Canadians (left panels) and the whole
population (right panels). The second row shows similar patterns in
the nation-wide General Social Survey. While the Aboriginal ESC is
a distinct sample from the General ESC, the panel labeled \textquotedblleft Aboriginal
GSS\textquotedblright{} is simply a subset of the full GSS sample.\label{fig:indigenous-distributions}}
\end{figure*}

This case study relates to the importance of being able to use life
satisfaction data across diverse cultural and economic circumstances.
It also demonstrates the use of the mixture model on a small sample.
The data come from two Canadian surveys: the national Equality, Security
and Community survey (General ESC, $N=3725$) and its follow-up small
sample of on- (70\%) and off-reserve (30\%) First Nations and Métis
peoples\footnote{Both of these groups are considered Aboriginal (and, along with the
Inuit, Indigenous).} in the Canadian Prairies (Aboriginal ESC, $N=446$). As can be seen
in the first panel of \figref{indigenous-distributions}, an enhancement
at \SWL=10 in the Aboriginal ESC sample makes it the modal response
value and may go some way to explaining the high mean reported \SWL.
Indeed, this is likely the first report of a \SWL distribution with
such a dominant response at its top value. On the other hand, respondents
also gave plenty of 7s, 8s, and 9s, each with higher frequency than
\SWL=5. Below I use the cognitive mixture model to assess how much
this distribution might be biased by \FVR, and whether the anomalous
estimates in \citet{Barrington-Leigh-Sloman-IIPJ2016-aboriginal}
are reversed.

The first column of \tabref{ESCaboriginal} shows a conventional ordered
logit estimate of 10-point life satisfaction of the Aboriginal sample.
For consistency with \citet{Barrington-Leigh-Sloman-IIPJ2016-aboriginal},
the education variable is a more continuous variable than in the previous
two applications, being measured on ten steps ranging from no primary
school to a PhD or professional degree. Once again, and despite a
sample size of only 446, a significantly negative coefficient on education
shows that, after adjusting for income, those with higher education
report \emph{lower} life satisfaction. In addition, the coefficient
on log household income is estimated to be most likely \emph{negative},
with a 95\% confidence interval between $-$0.40 and $+0.08$.

The second column reports the estimate of a cognitive mixture model.
Education strongly predicts numeracy, i.e., use of the full response
scale. Most importantly, the education anomaly in the earlier analysis
is resolved when \FVR is taken into account: the confidently negative
education coefficient is replaced by a weakly positive point estimate
with a 95\% confidence interval between $-$.08 and +.17. The weaker
anomaly of a negative income coefficient is also partly resolved;
in its place is one centered closely on zero with similar precision.

In order to address the surprisingly high average life satisfaction
reported by Indigenous respondents, I next use a pooled model to compare
groups after controlling for income and education. Pooled estimates
of the Canada-wide respondents and the First Nations/Métis sample
are shown in Columns (3) and (4) of \tabref{ESCaboriginal}. Adjusting
for income and education, the Aboriginal respondents report 0.30 higher
life satisfaction than non-Aboriginal. Although the explanatory variables
here are few and the model is simple, this positive boost is counterintuitive
for the reasons described above. However, when \FVR is accounted for
(Column 4), this situation is reversed, with the Aboriginal respondents
reporting a weakly lower life satisfaction than others with similar
income and education. In this model, education, income, and Aboriginal
status are all allowed to predict \FVR behavior. Education and income
positively predict lower propensity for \FVR behavior, as expected,
while Aboriginal status has the equivalent effect on \FVR as a two-point
reduction in educational attainment level, for instance from completing
a technical or community college certification to completing only
high school.

For the pooled sample, the mixture model estimates a 70\% higher
income coefficient and corrects the strongly negative education effect
of the naive model estimate with a weakly positive one.

\begin{table*}
  \centering
    \ifusetabularsource
      
    \renewcommand{\ctNtabCols}{5}
    \renewcommand{\ctFirstHeader}{\hline
 & \multicolumn{2}{c|}{Aboriginal sample} & \multicolumn{2}{c|}{Combined sample}\\ 

 & Conventional & Mixture model & Conventional & Mixture model\\ 

 & ologit & Flexible & ologit & Flexible\\ 

 & (1) & (2) & (3) & (4)\\ 
 \hline\hline
 }
    \renewcommand{\ctSubsequentHeaders}{\ctFirstHeader}
    \renewcommand{\ctBody}{Life satisfaction ($\beta_S$)&&&&\\  
&&&&\\  
\multicolumn{1}{|r|}{~ education (10 pt)}&\wrapSigFivePercent{$-$.082}&.044&\wrapSigOneThousandth{$-$.067}&.016\\  
&\coefse{.041}&\coefse{.064}&\coefse{.014}&\coefse{.019}\\  
\multicolumn{1}{|r|}{~ log(HH income)}&$-$.16&.001&\wrapSigOneThousandth{.32}&\wrapSigOneThousandth{.56}\\  
&\coefse{.12}&\coefse{.15}&\coefse{.046}&\coefse{.052}\\  
\multicolumn{1}{|r|}{~ Aboriginal sample}&&&\wrapSigFivePercent{.30}&$-$.11\\  
&&&\coefse{.13}&\coefse{.12}\\  
\hline Numeracy ($\beta_N$)&&&&\\  
&&&&\\  
\multicolumn{1}{|r|}{ constant}&&\wrapSigOnePercent{.77}&&\wrapSigOneThousandth{2.1}\\  
&&\coefse{.43}&&\coefse{.21}\\  
\multicolumn{1}{|r|}{ education (10 pt)}&&\wrapSigOneThousandth{.35}&&\wrapSigOneThousandth{.35}\\  
&&\coefse{.15}&&\coefse{.056}\\  
\multicolumn{1}{|r|}{ log(HH income)}&&&&\wrapSigOneThousandth{.54}\\  
&&&&\coefse{.12}\\  
\multicolumn{1}{|r|}{ Aboriginal sample}&&&&\wrapSigOneThousandth{$-$.74}\\  
&&&&\coefse{.15}\\
\hline FVRI&&\wrapSigOneThousandth{.34}&&\wrapSigOneThousandth{.15}\\  
&&\coefse{.076}&&\coefse{.017}\\    
\hline&&&&\\  
obs.&446&446&4171&4171\\  
log likelihood&$-$806&$-$821&$-$7409&$-$7307\\ 
\cline{1-\ctNtabCols}
 }
    
    \renewcommand{\ctCaption}{ {\color{blue} ~\\\begin{tabular}{lllrllrr}
\toprule
{} &   subsample &   model &     N &  ologit\-pid & mixture\-pid &  ologit\-R &  mixture\-R \\
\midrule
0  &   onreserve &  MODEL1 &   317 &  8933273224 &  7782519816 &       1.0 &       1.01 \\
1  &   onreserve &  MODEL2 &   317 &  3884732228 &  1483997528 &       1.0 &       1.01 \\
2  &  aboriginal &  MODEL1 &   446 &  9797339941 &  4416382936 &       1.0 &       1.00 \\
3  &  aboriginal &  MODEL2 &   446 &  5578514441 &  7841616665 &       1.0 &       1.03 \\
4  &     general &  MODEL1 &  3725 &  7186899617 &  8979126645 &       1.0 &       1.01 \\
5  &     general &  MODEL2 &  3725 &  3934267394 &  7682721639 &       1.0 &       1.01 \\
6  &    combined &  MODEL1 &  4171 &  4337765862 &  8823933257 &       1.0 &       1.00 \\
7  &    combined &  MODEL2 &  4171 &  4236994289 &  4749121579 &       1.0 &       1.00 \\
8  &    combined &  MODEL3 &  4171 &  8718257794 &  9136443414 &       1.0 &       1.01 \\
9  &    combined &  MODEL5 &  4171 &  6287817499 &  6915742277 &       1.0 &       1.00 \\
10 &    combined &  MODEL6 &  4171 &  3385252153 &  5583884131 &       1.0 &       1.01 \\
11 &    combined &  MODEL7 &  4171 &  5431648186 &  3725114946 &       1.0 &       1.00 \\
\bottomrule
\end{tabular}
 {\footnotesize\cpblColourLegend} } }
    \ifx\@ctUsingWrapper\@empty
    \begin{table}
    \begin{tabular}{|>{\bfseries}l|c|c|c|c|}
    \ctFirstHeader
    \ctBody
    \end{tabular}
    \end{table}
    \else
    \fi

    \renewcommand{\ctStartTabular}{\begin{tabular}{|>{\bfseries}l|c|c|c|c|}}
    \renewcommand{\ctStartLongtable}{\begin{longtable}[c]{|>{\bfseries}l|c|c|c|c|}}

            \ctStartTabular
            \ctFirstHeader
            \ctBody \hline
          \ctEndTabular%
      \else
        \includegraphics[scale=0.85]{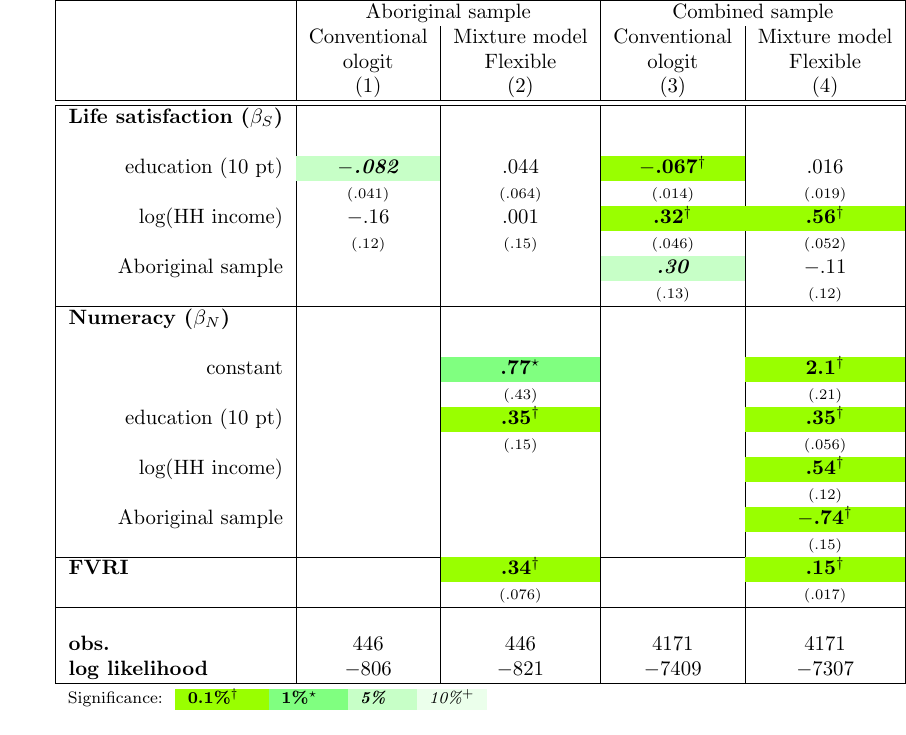}
        \fi \caption[Estimates of life satisfaction of Indigenous Canadians]{Estimates of life satisfaction of Indigenous Canadians\label{tab:ESCaboriginal}}
\end{table*}

\subsection{Ranking of U.S. states by happiness}

The United States is somewhat exceptional in that there are no prominent
domestic surveys assessing subjective wellbeing with more than a 4-point
response,\footnote{However, two international datasets, the World Values Survey and the
Gallup World Poll, do so on 10 and 11 point scales, respectively.} with the exception of the Gallup Daily Poll, which poses the Cantril
Ladder question on an 11-point scale.

In this section, I investigate the extent to which a ranking of states
by average reported life evaluations is biased by focal-value response
behavior. I find that state-level differences in educational attainment
relate to state-level differences in \FVR. Applying the cognitive
mixture model to these data provides a counterfactual ``latent''
or ``corrected'' mean life evaluation for each state, allowing for
a comparison between a naive ranking and a corrected ranking of states.

Ranking of happiness around the world garners considerable attention,
with over one million visits and downloads of the World Happiness
Report each year.  Below, the USA case demonstrates that a bias in
rankings occurs when mean responses are taken at face value.

\begin{figure*}
\begin{centering}
{\tabcolsep=3pt %
\begin{tabular}{cc}
\begin{tabular}{c}
\includegraphics[width=0.5\textwidth]{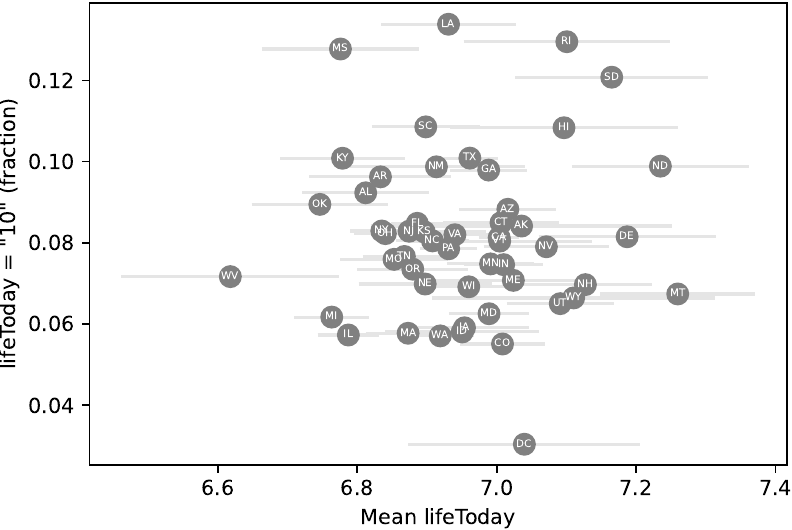}\tabularnewline
(a)\tabularnewline
\tabularnewline
(c)\tabularnewline
\end{tabular} & %
\begin{tabular}{c}
\includegraphics[width=0.5\textwidth]{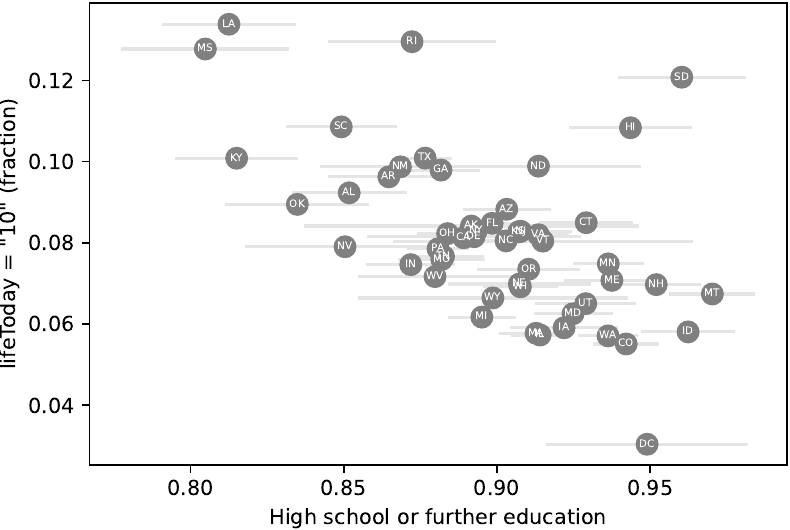}\tabularnewline
(b)\tabularnewline
\tabularnewline
(d)\tabularnewline
\end{tabular}\tabularnewline
\end{tabular}}\\
\includegraphics[width=0.7\textwidth]{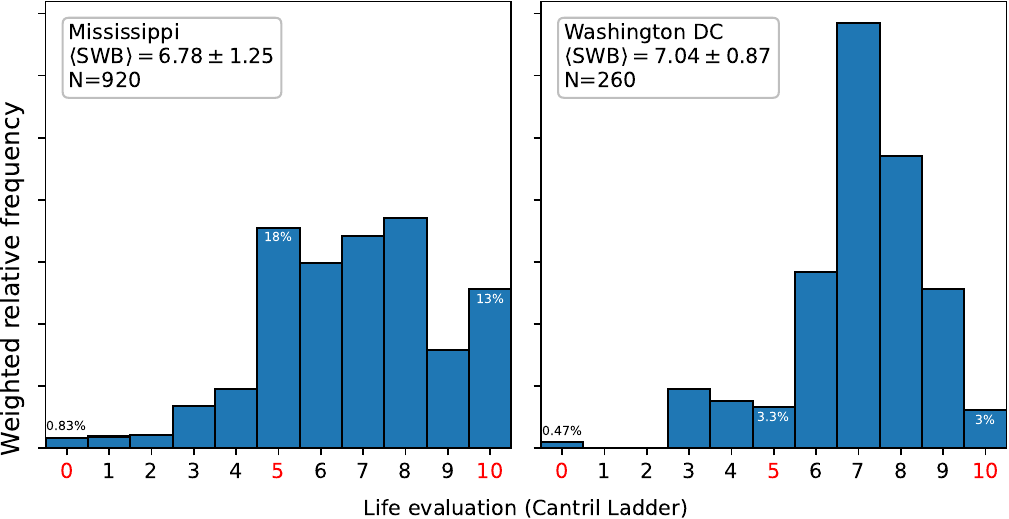}
\par\end{centering}
\caption[State-level responses and propensity to answer ``10'']{State-level responses and propensity to answer ``10'' . Use of focal
value \textquotedblleft 10\textquotedblright{} is uncorrelated with
mean life evaluation (a) but is negatively correlated with state average
education (b). Whiskers show standard errors. Panels (c) and (d) show
two examples of distributions with and without heavy \FVR behavior.
See \href{https://alum.mit.edu/www/cpbl/publications/Barrington-Leigh-JPubE2024-appendix.pdf}{Appendix Table F.6} for further descriptive statistics.
\label{fig:dailyPoll-isTen}}
\end{figure*}

\figref{dailyPoll-isTen}(a)'s horizontal axis shows the distribution
of state mean responses to the Cantril Ladder framing of life evaluation\footnote{See \href{https://alum.mit.edu/www/cpbl/publications/Barrington-Leigh-JPubE2024-appendix.pdf}{Appendix G.7} for the precise wording of the
question.} in the 2019 (final) wave of the Gallup Daily Poll. Counter-intuitively,
these means are uncorrelated with the fraction of respondents in each
state who provided the answer ``10'' on the 0--10 scale (vertical
axis). \figref{dailyPoll-isTen}(b) gives some suggestion as to why.
States with higher high school completion rates have lower tendency
to answer ``10''. \figref{dailyPoll-isTen}(c) and (d) show an example
of how much states can differ in terms of \FVR. The weighted response
distribution for Misssissippi, which has a high incidence of answer
``10'', is remarkably different from that of Washington DC,\footnote{The survey also covered Washington DC. It is aggregated here alongside
the states, even though it is entirely urban, unlike any state.} with the lowest incidence, even though their mean responses are similar.

With this motivation, \href{https://alum.mit.edu/www/cpbl/publications/Barrington-Leigh-JPubE2024-appendix.pdf}{Appendix Table F.7}
presents estimates of a version of the mixture model \href{https://alum.mit.edu/www/cpbl/publications/Barrington-Leigh-JPubE2024-appendix.pdf}{Appendix Equation A.1}
explaining individual responses with $\boldsymbol{x}=\boldsymbol{z}$
comprised of the logarithm of household income, along with a set of
indicators for a five-level educational attainment question. As before,
several parameters and posteriors of interest are: the fraction (\FVRI)
of respondents estimated to be using a simplified focal value scale;
a mean of the latent life evaluation which would have been observed
had all respondents chosen to use the full scale; and coefficients
for the effect of income and education levels on the underlying (latent)
life evaluations. These values are estimated separately for each state
and can be compared in \href{https://alum.mit.edu/www/cpbl/publications/Barrington-Leigh-JPubE2024-appendix.pdf}{Appendix Table F.7}
to the naive model, equivalent to an ordered logit, in which focal
value behavior is not acknowledged.

\begin{figure*}
\centering{}\includegraphics[width=1\textwidth]{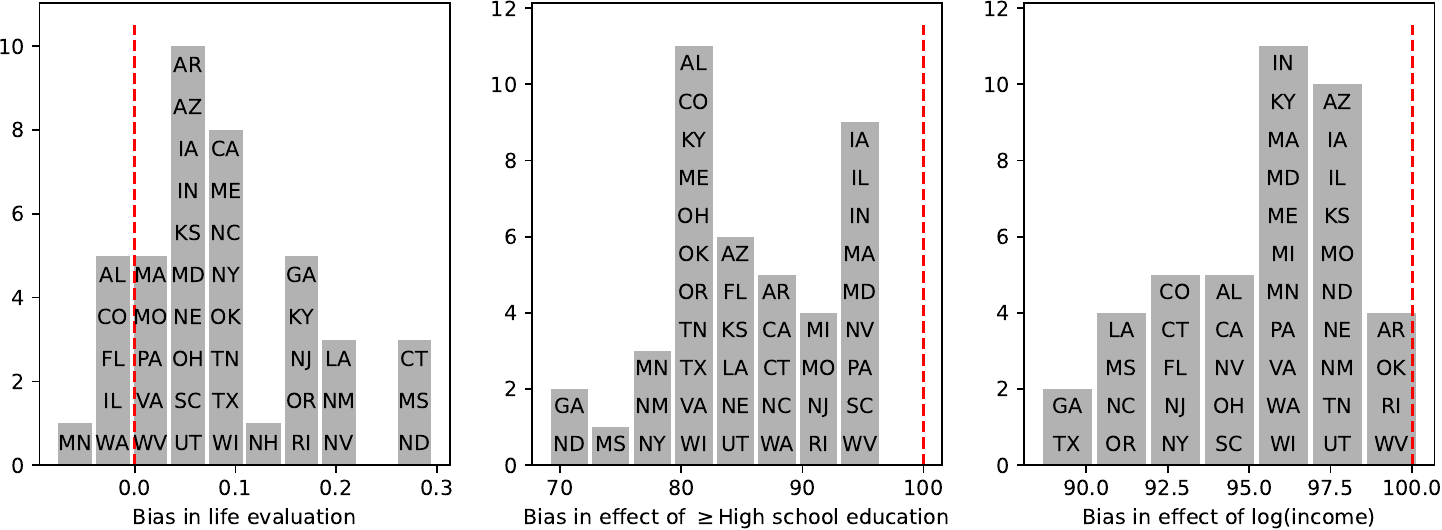}\caption[Distributions of bias by state in mean SWB and estimated effects]{Distribution across states of bias in mean life evaluations (in units
of the 0--10 scale) and in estimated effects (in odds ratios as a percentage)
on life evaluations. For instance, an odds ratio of 95\% in the log(income)
plot means that the bias results in a 5\% reduction in the probability
of being one level higher on the 0--10 scale, all else equal, in response
to a unit increase in log(income)..\label{fig:histograms-three-biases-dailypoll}}
\end{figure*}

\figref{histograms-three-biases-dailypoll} presents the distributions
of biases in mean life evaluation and effects of high school completion
and family income on life evaluations, obtained by comparing the ordered
logit and mixture models. It shows that the Cantril Ladder question
is in most states estimated to elicit highly positively-biased responses.
In other words, the effect of ``rounding up'' to 10 (or to 5) outweighs
any rounding down to 5 (or to 0), and is large. In many cases, the
raw mean report is 0.1--0.2 higher than that inferred with the focal
value correction, which is large given that the standard deviation
of Cantril ladder means is 0.13 among states, and the standard deviation
of individual responses nationally is only 1.89. This bias is larger
for states with lower educational attainment.

\figref{histograms-three-biases-dailypoll} also shows that the distributions
of biases in education effects and in income effects are both uniformly
downwards at the state level. Reassuringly, the mixture-model estimated
effects of educational attainment on wellbeing are overwhelmingly
positive after the correction (\href{https://alum.mit.edu/www/cpbl/publications/Barrington-Leigh-JPubE2024-appendix.pdf}{Appendix Table F.7}).

\begin{figure*}
\begin{centering}
\includegraphics[width=1\textwidth]{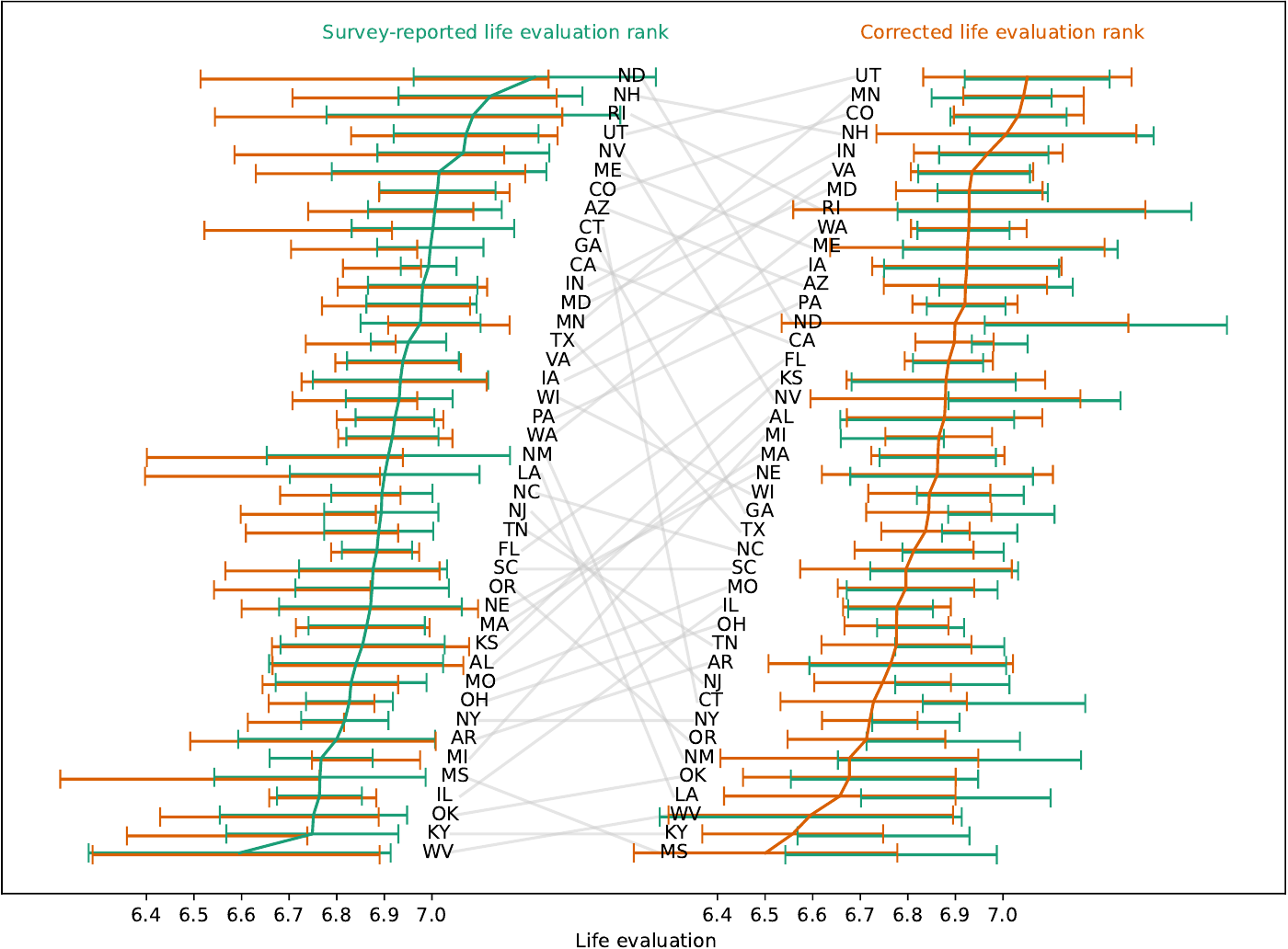}
\par\end{centering}
\caption[Observed and corrected U.S. state rankings]{Observed and corrected U.S. state rankings. Error bars show 95\%
confidence intervals. States without at least one of each possible
response to the \SWB question are omitted.\label{fig:daily-poll-ranks}}
\end{figure*}

Lastly, \figref{daily-poll-ranks} presents state rankings for both
the raw reported life evaluation and the estimated latent life evaluation.
The overlapping estimate ranges reflect the typically imprecise nature
of this kind of ranking, especially given the small sample size in
some states (see \href{https://alum.mit.edu/www/cpbl/publications/Barrington-Leigh-JPubE2024-appendix.pdf}{Appendix Table F.6}). There is also
significant consistency (correlation 0.70) between the corrected and
uncorrected rankings. Nevertheless, the shifts are considerable:
more than a quarter of states shift by more than a quartile in the
distribution (despite the overall correlation), 65\% of states shift
positions by 5 or more, and 37\% shift by 10 or more. 

\section{Discussion and conclusion\label{sec:Conclusion}}

The contributions of this paper are to \begin{inparaenum}[(i)] \item 
explain a prominent feature of many subjective scale response distributions
as the result of respondents simplifying the scale; \item  identify
education and other proxies of numeracy as predictors of this ``focal
value rounding'' behavior; \item  formulate a model and estimation
strategy for predicting life satisfaction responses from individual
and contextual circumstances which properly takes into account a mixture
of reporting behavior used by respondents; \item  explore theoretically
the biases possible due to the effect;  \item  provide a way to
estimate the degree (\FVRI) of focal value rounding behavior; and \item 
demonstrate the application of the estimation method and its significant
impact for four published studies and surveys.\end{inparaenum}

\citet{Clark-Oswald-JPubE1996} write ``Counter to what neoclassical
economic theory might lead one to expect, highly educated people appear
less content. The effect is monotonic and well-defined''. This contradiction
with neoclassical economic theory has generally held up to subsequent
analysis over two decades but is partly resolved with the model described
here, which takes into account a conspicuous empirical feature of
the subjective wellbeing response function.

Income effects have been a focus in the study of wellbeing in economics
since the field's inception, and an enormous literature exists around
the magnitude of the income coefficient \citep[e.g., ][]{Easterlin-1974,Deaton-JEP2008-GWP,Clark-Frijters-Shields-JEL2008,Dolan-Peasgood-White-JEPsych2008,Easterlin-JEBO1995,Easterlin-EI2013-SWB-growth-public-policy,Ferrer-i-Carbonell-JPubE2005-veblen-GSOEP,Kapteyn-vanPraag-vanHerwaarden-EL1978,Luttmer-QJE2005,Senik-JES2005,VanPraag-EER-1973fei}.
Almost every economic study of \SWL includes an estimate of the income
effect, and typically other influences on life satisfaction are quantified
in terms of their income ``compensating differential'', i.e., the
ratio between a coefficient of interest and the coefficient on income.
Thus, the large corrections estimated here for the income coefficient
indicate that material supports are slightly more effective for raising
human wellbeing, as compared with the other --- especially social ---
dimensions of life, than the literature has shown so far. According
to the simulations, some downward bias can also occur for those other
coefficients, especially when those dimensions of life are correlated
with education, but there is little empirical evidence for this in
the estimates carried out in this paper.

One next step for research is to examine international and cultural
patterns in response functions.  Effects will differ across countries
according to where the average \SWL level lies on the scale, and according
to the income and education distribution. There may be additional
international differences in the tendency to use focal values. Therefore,
using the mixture model approach, both differences in education systems
and more cultural drivers of \FVR can be incorporated into international
comparisons of \SWL. Flexibly modeling each possible \SWL response so
as to allow for non-ordinal relationships between them, carried out
here using multinomial logit, is a good starting point for detecting
such response biases driven by cultural norms as well as numeracy.
 Despite the general evidence of good comparability of \SWL patterns
across cultures \citep{Helliwell-Barrington-Leigh-Harris-Huang-2010},
it may still be possible to identify response biases towards central
values or away from ``extreme'' values. One natural extension of
the model described in this paper is to allow for the inclination
to round (\FVR) to vary separately for each focal value, effectively
creating a mixture of eight ``types'' in the case of three focal
values.

A deeper analysis of panel data will also be important, through an
extension to incorporate fixed effects into the model developed in
this paper. Preliminary analysis of panel data with a 5-point scale
for \SWL, treating values 1, 3, and 5 as focal values, shows that the
probability of \SWL changing from the middle value is decreasing in
education. Traditional 1st-differences approaches for panel fixed
effects are invalid because, for instance, the dependence of the $3\rightarrow4$
transition is not the mirror of the $4\rightarrow3$ transition.

Another extension of the model used in this paper will be to incorporate
instrumental variables. Fortunately, this is relatively straightforward
in Bayesian estimation frameworks, in which a single-step estimation
procedure for instrumental variables is natural, subject to the normal
exclusion restrictions \citep{Dreze-Econometrica1976-Bayesian-Instrumental-Variables,Kleibergen-Zivot-JEconometrics2003-Bayesian-Instrumental-Variables}.

As a proof of principle and in light of the descriptive evidence,
this paper focuses on the idea of numeracy and on education as a primary
predictor of \FVR. Understanding the role of secondary influences,
such as other demographic variables, fatigue, the cost of time, or
motivation with respect to the survey, may help to identify other
biases or to design better surveys.

Survey and questionnaire interface design is a further topic of future
work. While the present study carries out an \emph{ex post }determination
of how respondents have used a subjective numerical scale, it may
make sense to give respondents this choice up front. An interactive
survey interface could dynamically offer different degrees of precision
or resolution in responses, thus accommodating variation in cognitive
capacity and other differences in the confidence of respondents' answers.
 Open-ended graphical scales may be one means to accomplish this,
but further research into ways to elicit a statement of precision
from respondents would be valuable. The potential for creativity and
innovation is high, given the increasing availability of technology
during an interview.

Depending on one's perspective, the present findings on response behavior,
happiness income coefficients, and mean response biases may be taken
as a warning of how difficult it would be to realize the most ambitious
implementations of \SWL as a guide to policy \citep{Frijters-Clark-Krekel-Layard-BPP2020-Happy-Choice-SWB-as-goal-for-government,Frijters-Krekel-2021-SWB-policy-handbook,Barrington-Leigh-Escande-SIR2017-review-indicators,Barrington-Leigh-2016-CICbook-SWB-community-indicators,happiness-research-institute-2020-WALYs,Barrington-Leigh-SNSS2021-budgeting-for-happiness,UK-Treasury-2021-GreenBook-supplement-wellbeing,MacLennan-Stead-Rowlat-UK-Treasury-2021-life-satisfaction-approach}
or, conversely, as another reassuring example of the  robustness
of \SWL inference to potential flaws inherent in its cognitive complexity,
and possibly even a defense of the rough magnitudes of estimated effects
that have become so reproducible in study after study. I take away
both of these messages.

\section*{Acknowledgements}

I am grateful for discussion and comments from Fabian Lange, Kevin Lang, Andrew Oswald, Idrissa Ouili, Nadia DeLeon, two helpful referees, 
co-editor Keith Marzilli Ericson,
and audiences  in Vancouver, Chicago, Oxford, Cambridge, Montreal, and Winnipeg, among others. 
This work was supported by Canada's Social Sciences and Humanities Research Council (SSHRC) grant 435-2016-0531.



\bibliographystyle{elsarticle-harv}



\end{document}